\documentclass[useAMS,usenatbib]{mn2e}
\usepackage{graphicx}

\def\revised{}

\title[The contribution of starspots to coronal structure]{The contribution of starspots to coronal structure}
\author[D. Arzoumanian, M. Jardine, J.-F. Donati, J. Morin and C. Johnstone]{D. Arzoumanian$^{1,*}$, M. Jardine$^{1}$, J.-F. Donati$^{2}$, J. Morin$^{3}$ and C. Johnstone$^{1}$\\
$^{1}$School of Physics and Astronomy, University of St Andrews, St Andrews, Scotland KY16 9SS\\
$^{*}$Now at: CEA/DSM-CNRS-Universit\'e Paris Diderot, IRFU/Service dÕAstrophysique, CEA Saclay, Orme des Merisiers, 91191 Gif-sur-Yvette, France\\
$^{2}$Laboratoire d'Astrophysique, Observatoire Midi-Pyr\'en\'ees, F-31400 Toulouse, France\\
$^{3}$Dublin Institute for Advanced Studies, School of Cosmic Physics, 31 Fitzwilliam Place, Dublin 2, Ireland}

\begin{document}


\pagerange{\pageref{firstpage}--\pageref{lastpage}} \pubyear{2002}

\maketitle

\label{firstpage}

\begin{abstract}
Significant progress has been made recently in our understanding of the structure of stellar magnetic fields, thanks to advances in detection methods such as Zeeman-Doppler Imaging. The extrapolation of this surface magnetic field into the corona has provided 3D models of the coronal magnetic field and plasma. This method is sensitive mainly to the magnetic field in the bright regions of the stellar surface. The dark (spotted) regions are censored because the Zeeman signature there is suppressed. By modelling the magnetic field that might have been contained in these spots, we have studied the effect that this loss of information might have on our understanding of the coronal structure. As examples, we have chosen two stars (V374 peg and AB Dor) that have very different magnetograms and patterns of spot coverage.  We find that the effect of the spot field depends not only on the relative amount of flux in the spots, but also its distribution across the stellar surface. For a star such as AB Dor with a high spot coverage and a large polar spot, at its greatest effect the spot field may almost double the fraction of the flux that is open (hence decreasing the spindown time) while at the same time increasing the X-ray emission measure by two orders of magnitude and significantly affecting the X-ray rotational modulation.

\end{abstract}

\begin{keywords}
stars: magnetic fields $-$ stars: coronae $-$ stars: spots   $-$  stars: individual: V374 Peg, AB Dor $-$  stars: imaging $-$ techniques: polarimetric 
\end{keywords}

\section{Introduction}

The first detection of a magnetic field in a star, our Sun, was obtained  a century ago by \citep{Hale.1908}  thanks to Zeeman's discovery of the effect of magnetism on radiation 12 years before. Hale observed and correctly interpreted the magnetic polarisation of spectral lines
in sunspots and attributed it to magnetic fields of nearly 3~kG. Since this detection, the understanding of solar and stellar magnetic fields has improved,
thanks mostly to the improvement of instrumental performance and to the sophistication
of numerical simulations.

Magnetic fields are known to be present in a wide variety of stars, from very low-mass M dwarfs to super-massive O stars \citep{Donati.2009}. They play a role at basically all evolutionary stages, from collapsing molecular clouds and very young protostars to supernovae, degenerate white dwarfs and neutron stars. Magnetic fields are found to influence significantly a number of physical processes operating within and in the immediate vicinity of stars, such as accretion, diffusion, mass loss, turbulence and fundamental quantities such as mass, rotation rate and chemical composition.

Periodic changes in brightness  which indicate cooler or brighter starspots on the surface have long been observed in stars. In order to understand the magnetic field structure of a star it is not enough to know that spots exist, their location and extent are also important.

While active stars are too distant  to be resolved directly, indirect techniques such  as spectropolarimetry using Zeeman-Doppler Imaging (ZDI) have allowed the mapping  of stellar surface magnetic fields  \citep{Donati.Cameron.1997,Donati.1999}.
While ZDI  is the most successful of the  methods of field detection, {\revised  Doppler Imaging (DI)  aims to reconstruct the starspot distribution  which is contained in time varying line profiles of rotating stars \citep{Berdyugina.2005}. Both techniques  take advantage of the fast rotation and the inclination of some stars to reconstruct the surface magnetic field and the surface brightness using  Zeeman  and  Doppler effects.}

\begin{figure}
\resizebox{8.0cm}{!}{
\includegraphics{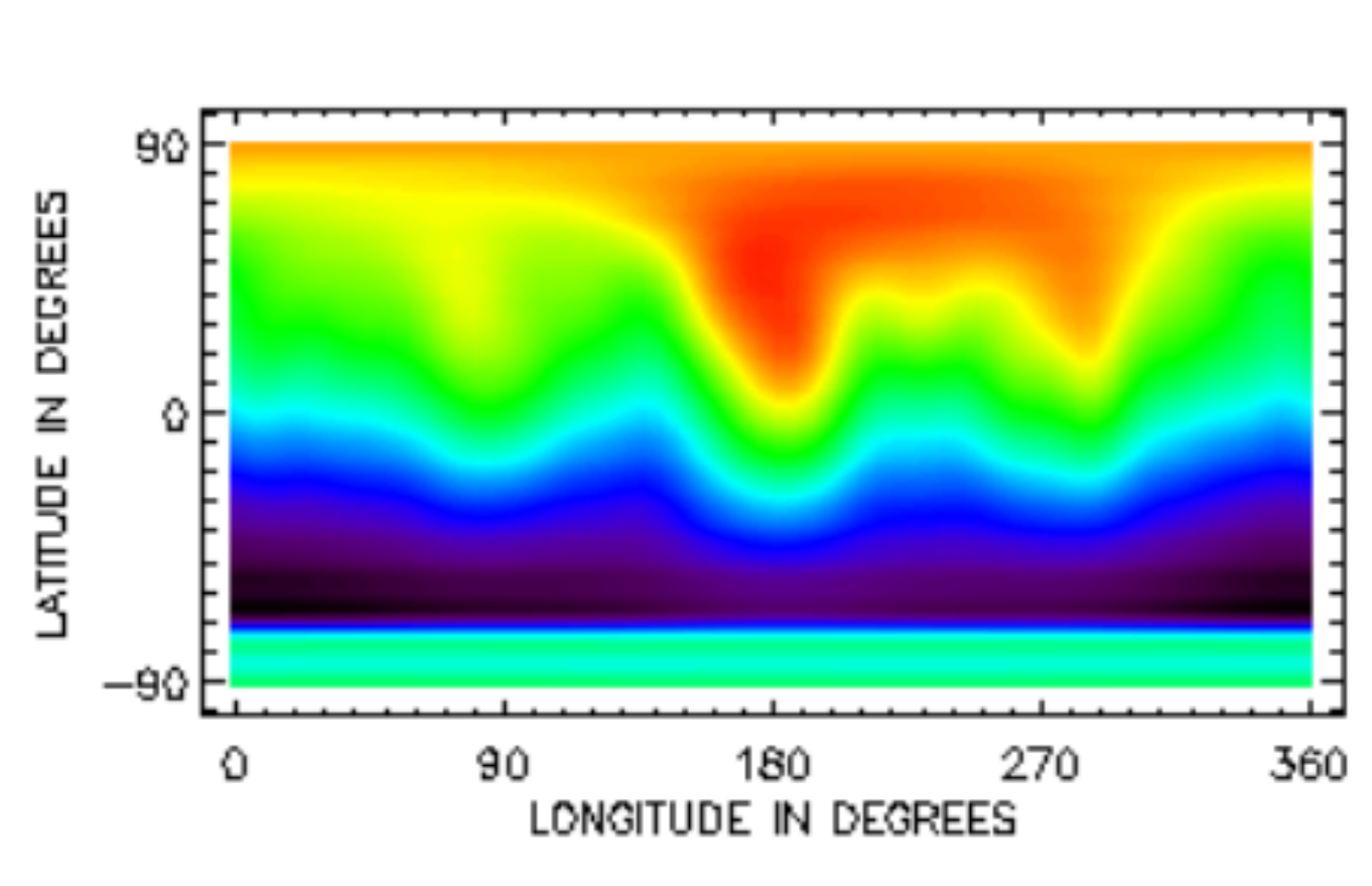}}
\resizebox{7.8cm}{!}{
\includegraphics{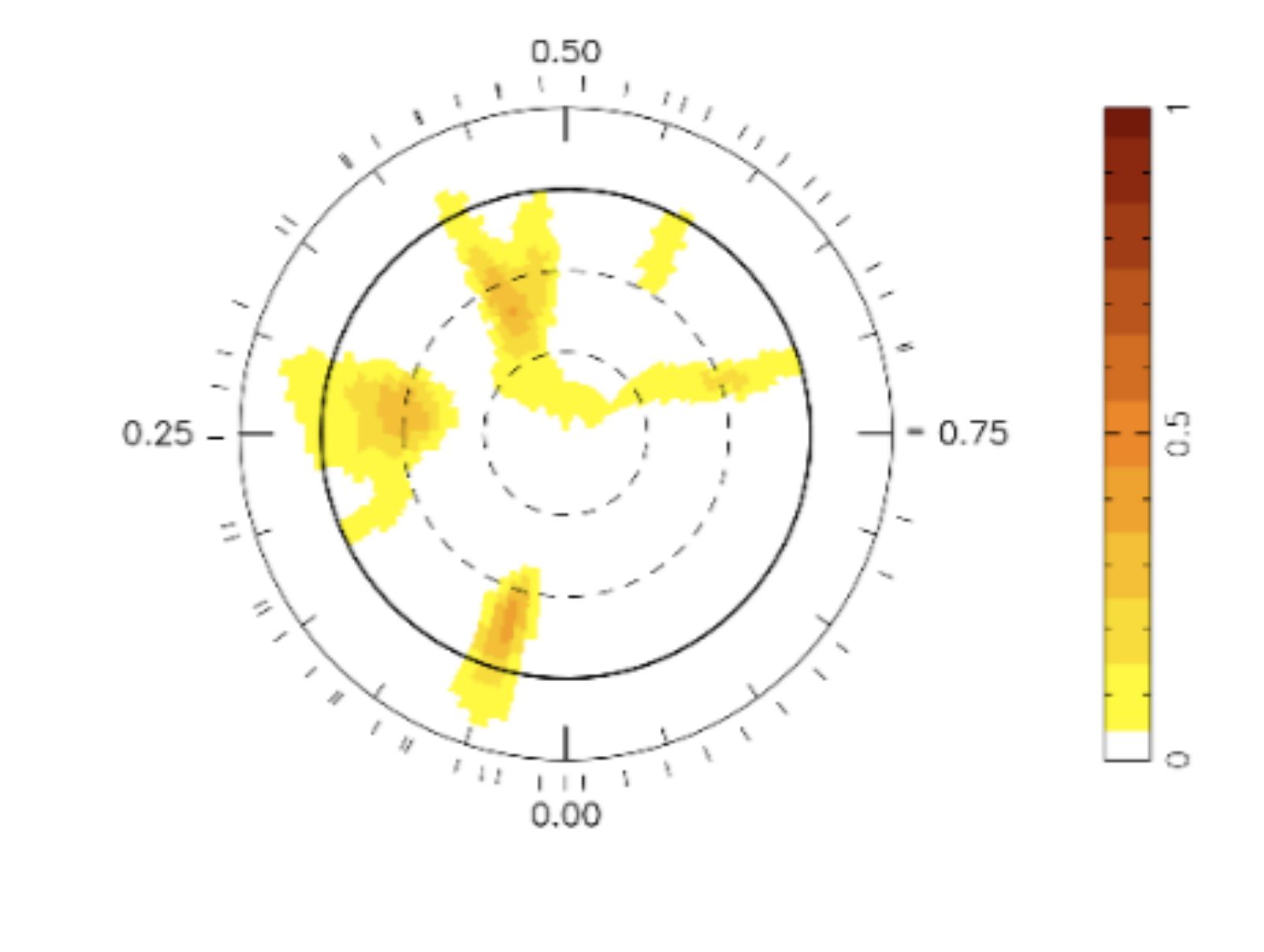}}
\resizebox{8.0cm}{!}{
\includegraphics{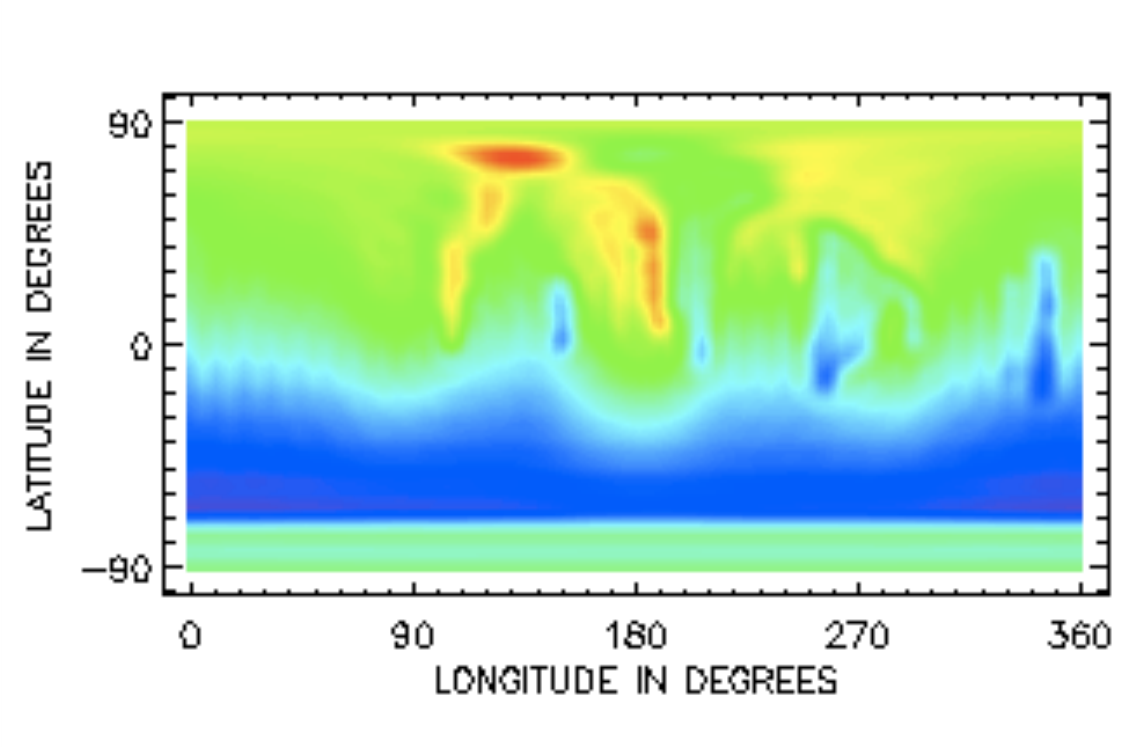}}
\caption{Top: magnetogram of V374 Peg from observations  in August 2005.  The colours represent the magnitude and polarity of the radial component of the field: red represents positive polarity, while blue represents negative polarity. The  maximum value of the surface magnetic field strength is 1.7~kG. There is limited information in the lower hemisphere because of the orientation of the star's rotation axis to the line of sight. Middle: brightness map of V374 Peg, the colour table shows spot filling factor. Bottom: Example of the new magnetogram of V374 Peg after the addition of the spot-map to the ZDI original map. The added spot map was created by contouring spots above the 0.05 level of brightness and assigning them a field strength of 1000~G.}
\label{0805disp} 
\end{figure}
   
\begin{figure}
\resizebox{8.5cm}{!}{
\includegraphics{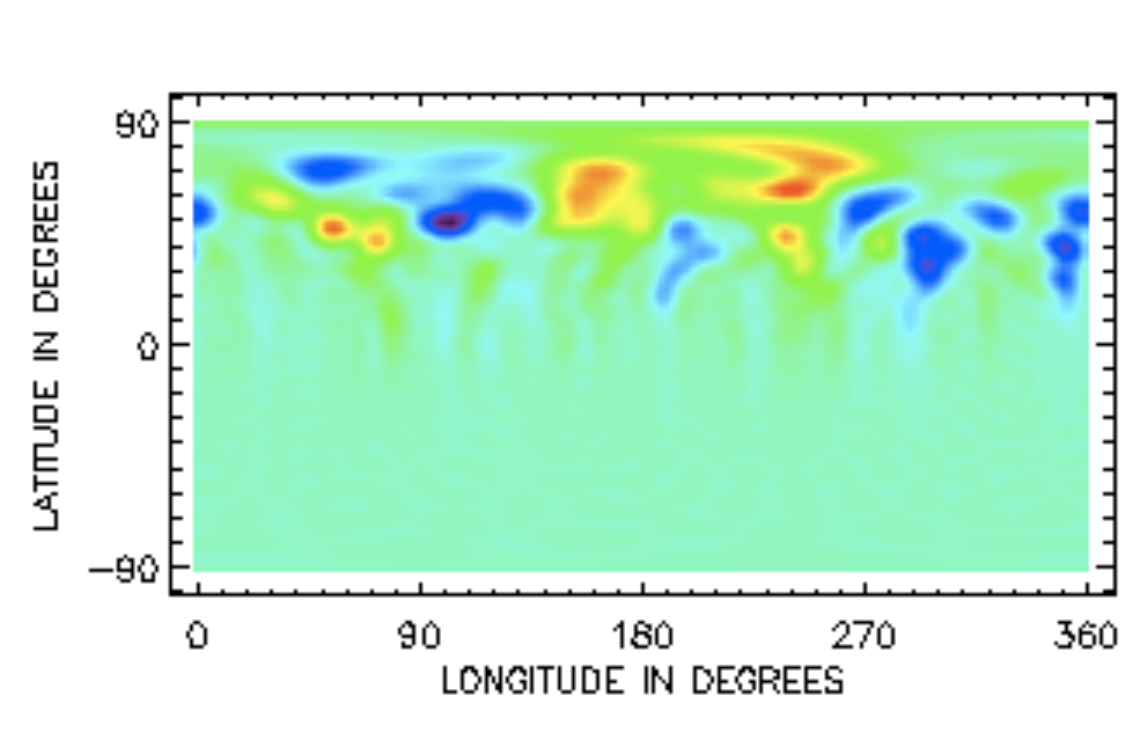}}
\resizebox{8.5cm}{!}{
\includegraphics{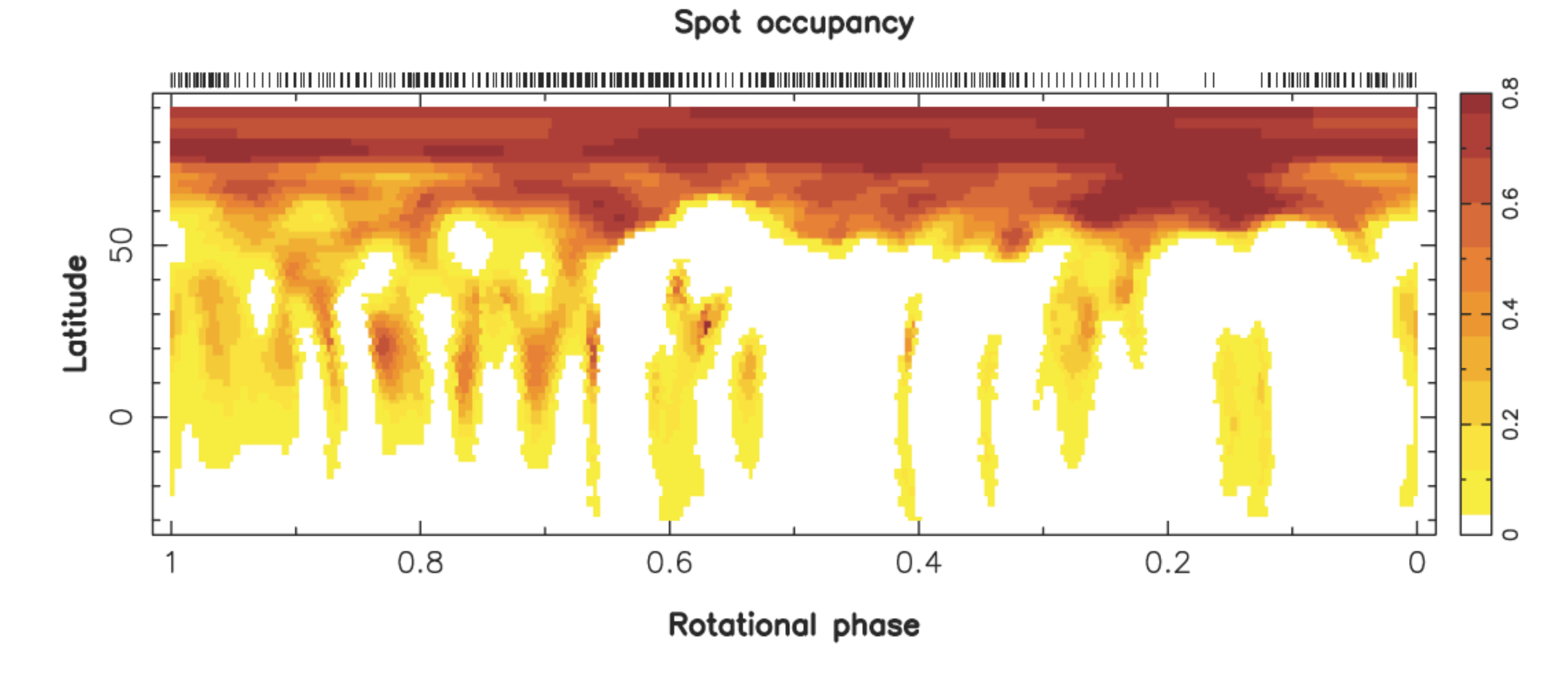}}
\resizebox{8.5cm}{!}{
\includegraphics{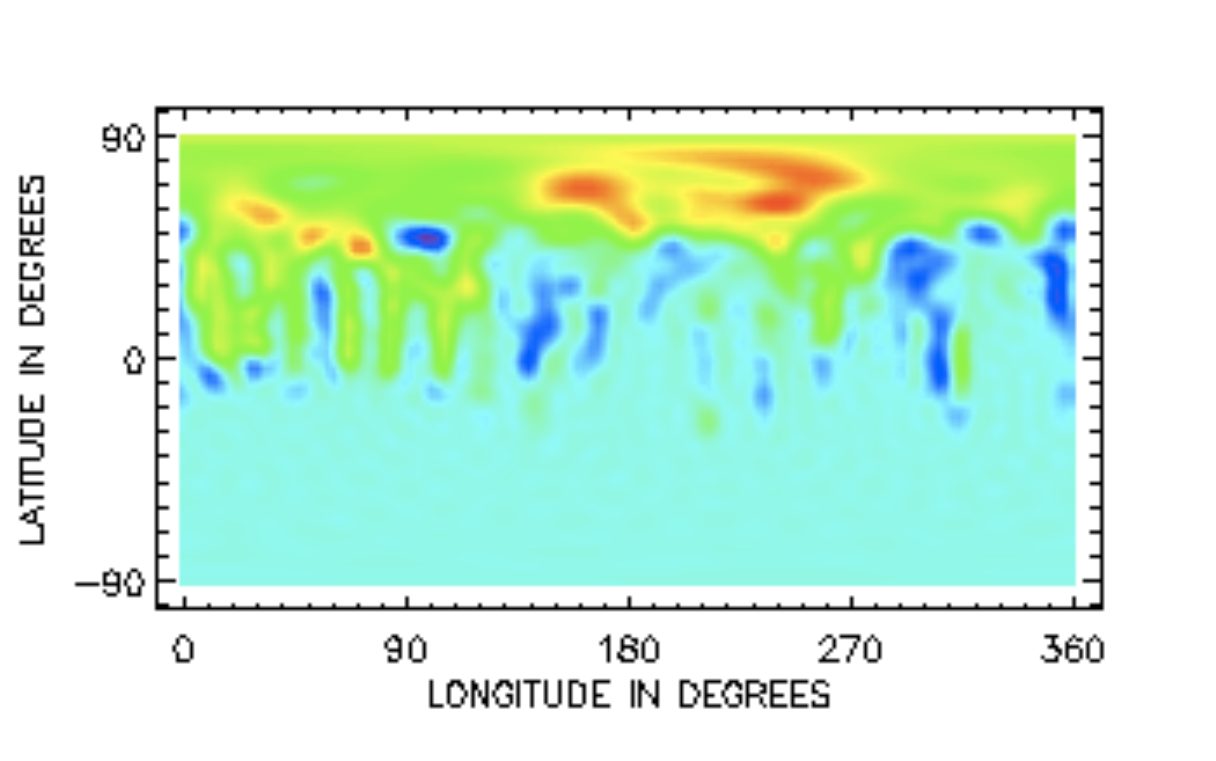}}
\caption{Top: magnetogram of AB Dor (observations carried on in  2004).  The red color represents  positive polarity , the blue  negative polarity. There  is limited information for latitudes below $60^{\circ}$ in the lower hemisphere because of the inclination of AB Dor's rotation axis  to the line of sight.  Middle: brightness map of AB Dor. Bottom: Example of a new magnetogram of AB Dor  after  addition of the spot-map to the ZDI original map. The added spot map was  created by contouring spots at levels between 0.05 and 0.1 with a field strength of 500~G and at levels over 0.1 with a field strength of 900~G (case 1).}
\label{ABdor}
\end{figure}


From the derived maps of the surface magnetic  fields of  stars we extrapolate the three-dimensional coronal field structure  using a {\em Potential Field Source Surface} extrapolation model developed by \citet{Altschuler.1969}. From the modeled stellar corona it is possible to calculate the fraction of the magnetic flux that is open (and hence contributing to the loss of mass and angular momentum in the stellar wind) and the tilt of the large-scale dipole component of the field. We can also derive the gas  pressure and the electron density at every location along the field lines, the mean electron density in the corona and the filling factor of the corona, as well as  the magnitude and rotational modulation of the X-ray emission measure. These calculated parameters can be compared to some observed values of the X-ray emission measures and coronal mean densities {\revised (cf. \citet{Hussain.2007} for AB Dor)}.

As stars are unresolved, the Doppler effect in the ZDI technique  is used to localize the emission coming from the stellar surface using the redshifted and blushifted wavelengths that appear during the rotation of the star. 
In order to differentiate  between magnetic features above and below the equator   the stellar rotation axis has to  be tilted to a certain extent to the line of sight. This means that  ZDI maps are unable to reproduce the field over the entire surface of any star. Some of the surface magnetic field may also be neglected because of the presence  of dark regions on the surface of the stars (dark spots) generally assumed to be the stellar analogue of sunspots  {\revised (sunspots have fields of  2 - 3~kG)}. If  this analogy is valid, then they could  contain strong magnetic fields which are {\revised almost missed by the spectropolarimetric observations. Indeed, the luminous flux emitted by dark spots being lower than in the quiet
photosphere contributes  little to the stellar spectra. A
simple estimate of the contrast between quiet photosphere and spots is the
ratio of the black-body fluxes at corresponding temperatures  taken at the
central wavelength $\lambda_0$ of the LSD line \citep{Donati.1997} used to
reconstruct ZDI maps. Using the contrast in temperature between the spots and the quiet photosphere
($\Delta T_{qs}$) from \citet{Berdyugina.2005}, we find flux ratios of 2.5 for V374
Peg ($\Delta T_{qs}=400$~K, $\lambda_0=700$~nm), and 3.7 for AB~Dor ($\Delta
T_{qs}=1000$~K, $\lambda_0=550$~nm).}

\begin{figure*}
\resizebox{18.5cm}{!}{
\includegraphics{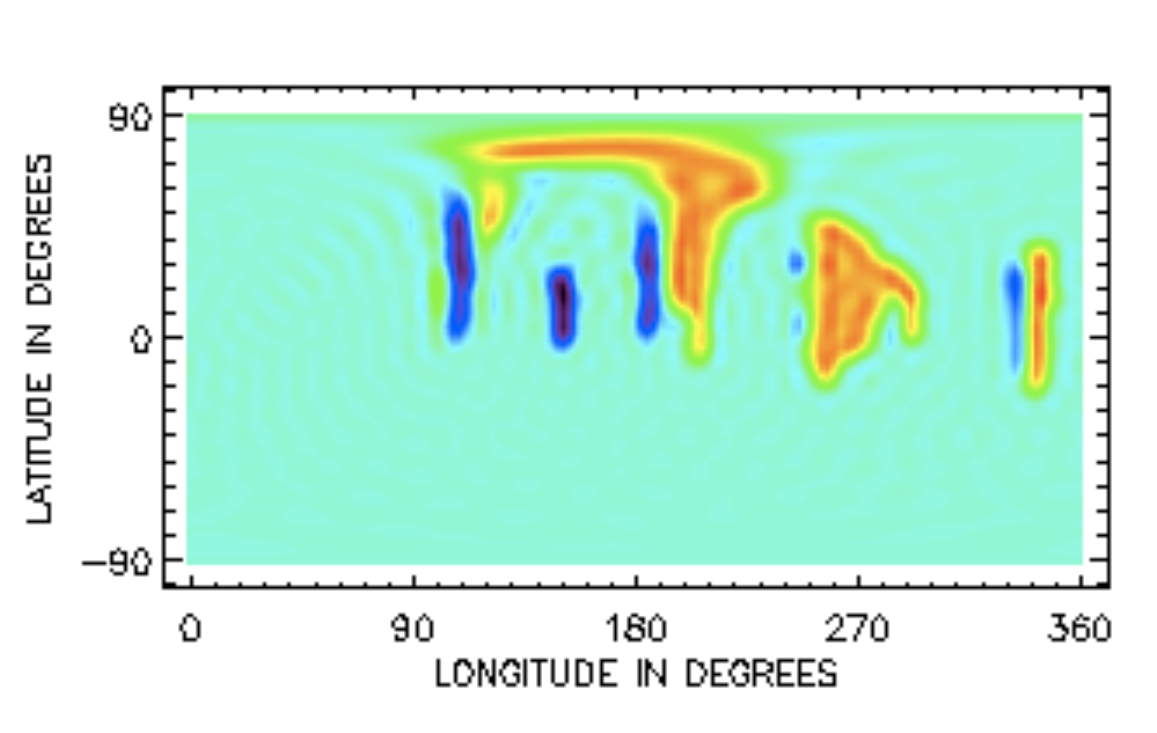}
\includegraphics{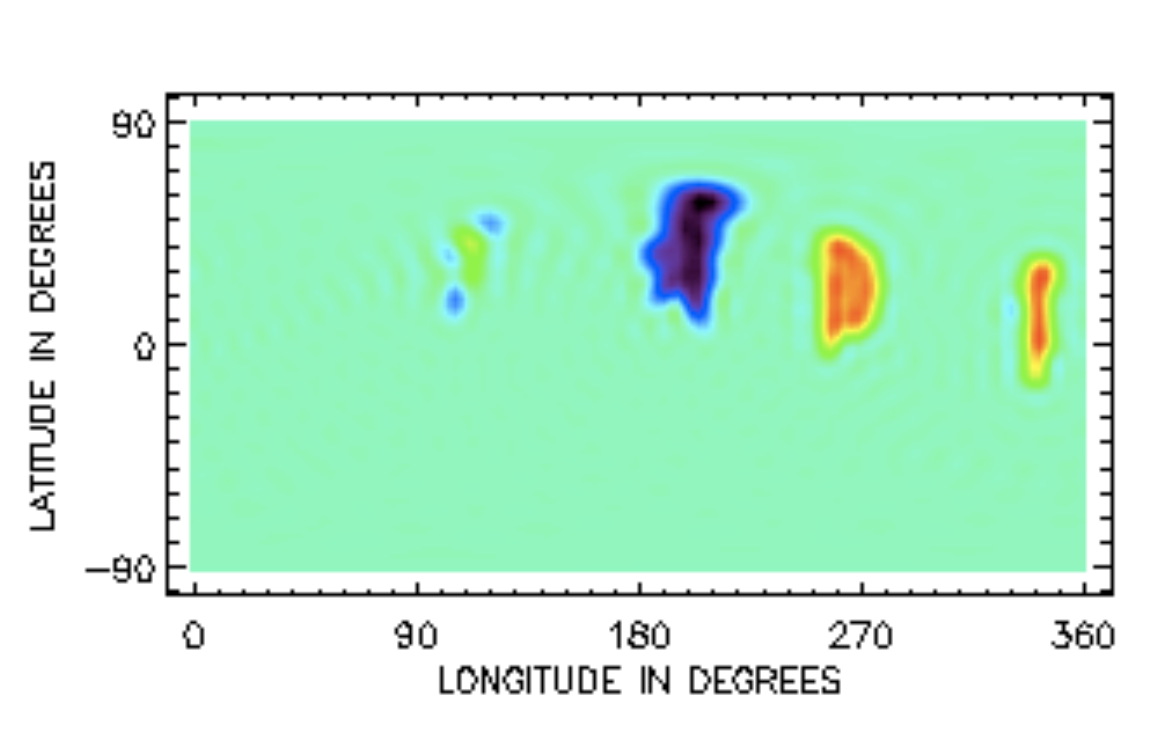}
\includegraphics{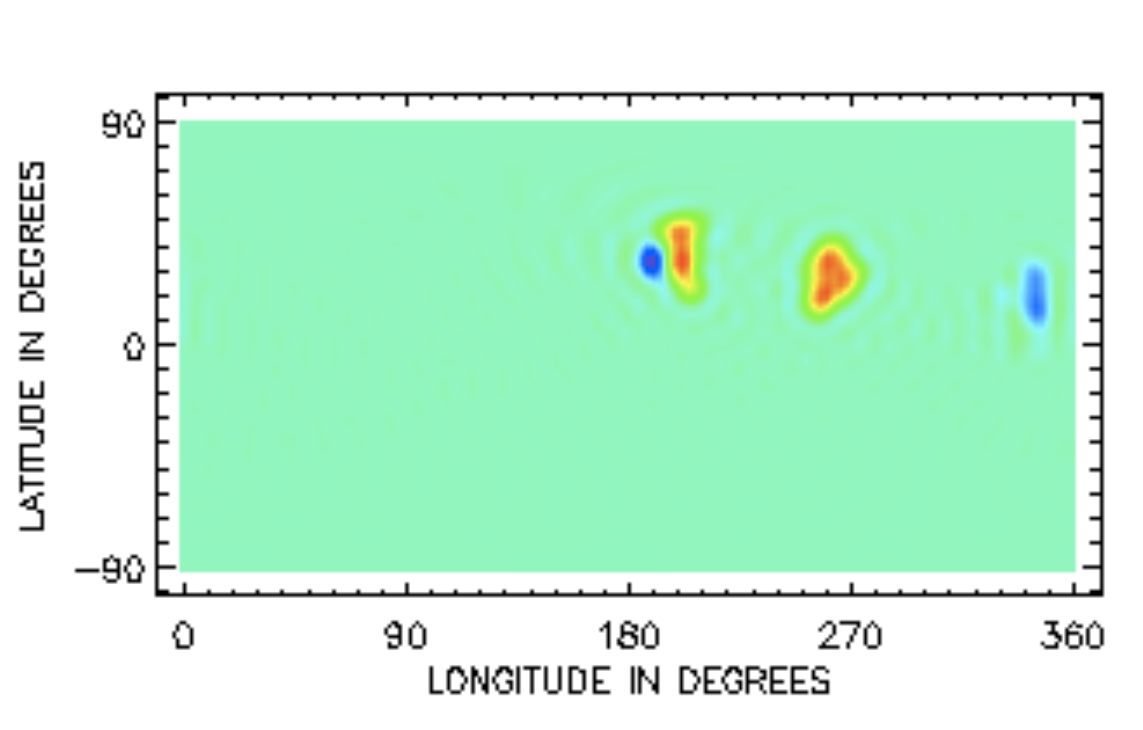}}
\caption{Examples of spot-maps derived from the brightness map of V374 Peg considering  spots with  brightness respectively above 0.05 on the left, 0.1 in the middle  and 0.2 on the right. The red represents a positive polarity, the blue  a negative polarity.}
\label{0805spots} 
\end{figure*}


\begin{figure*}
\resizebox{18.5cm}{!}{
\includegraphics{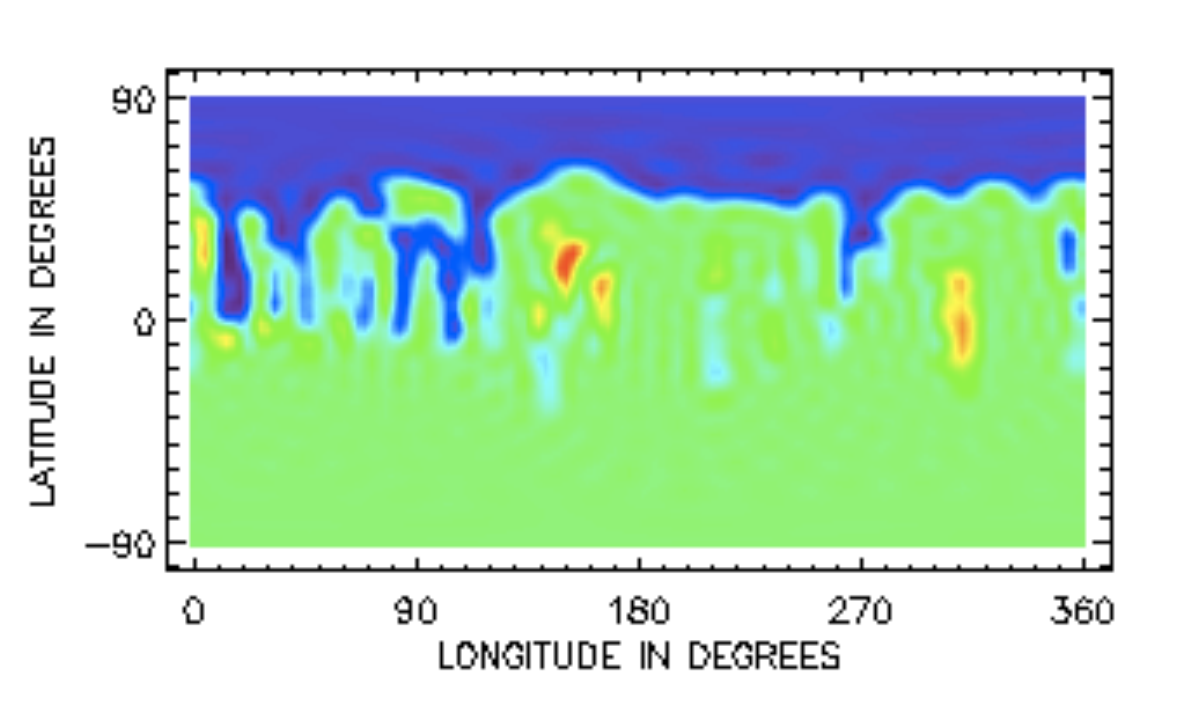}
\includegraphics{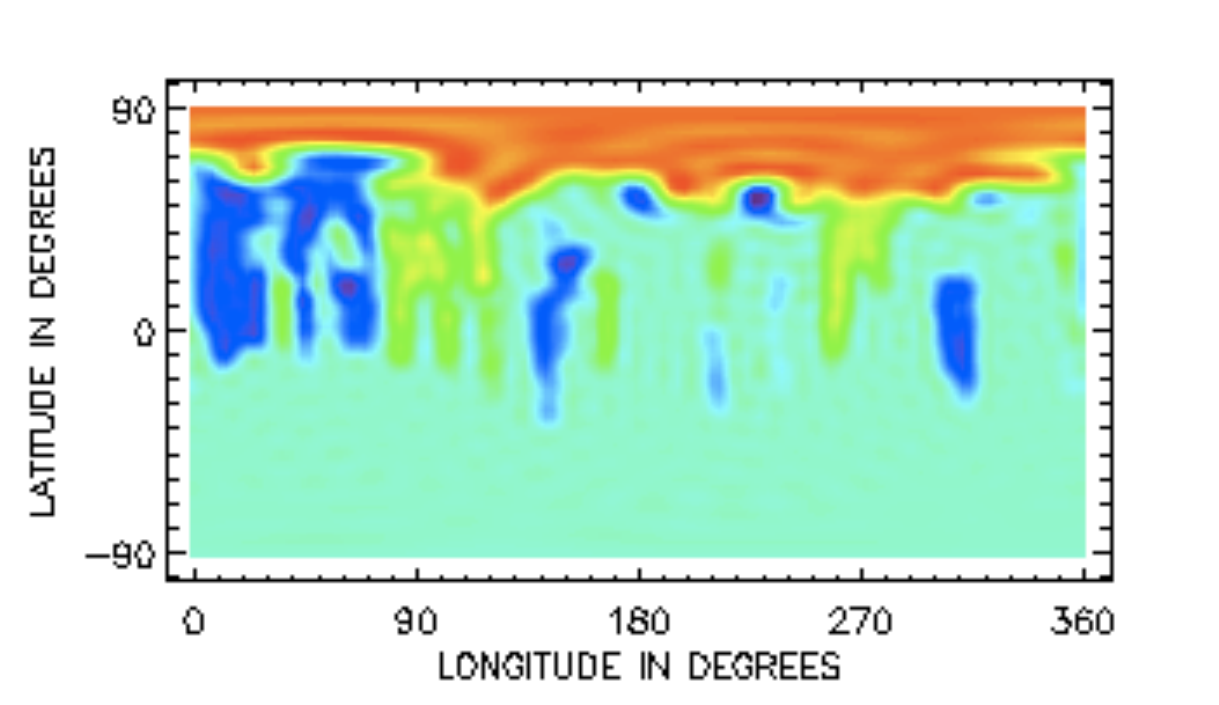}
\includegraphics{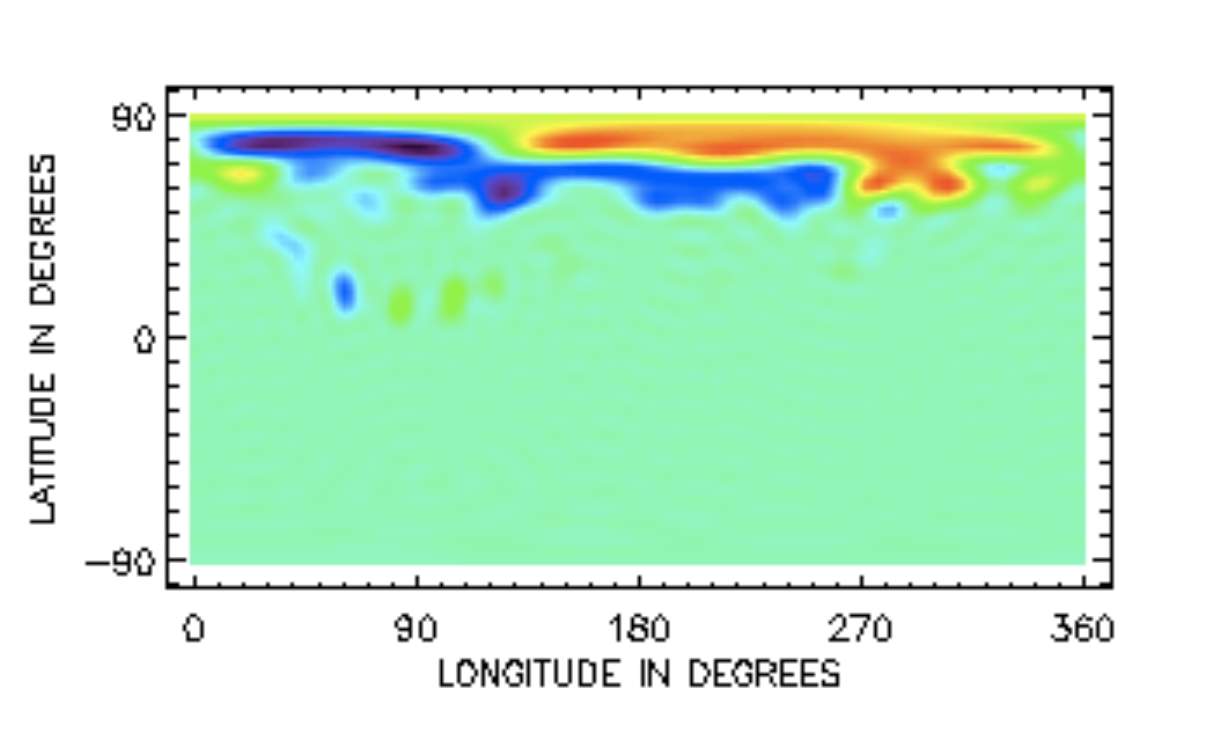}}
\caption{Examples of spot-maps derived from the brightness map of AB Dor considering the three different cases detailed in the text: case 1 on the left, case 2 in the middle and case 3 on the right. The red color represents a positive polarity, the blue   a negative polarity.}
\label{ABDor_spots}
\end{figure*}

The incompleteness of the ZDI maps comes also from the finite resolution of the observations.
As  the circular polarization from the longitudinal Zeeman effect is to first order
proportional to the net magnetic flux through the angular resolution element,
if the magnetic field has mixed-polarity fields inside the resolution element
with equal total amounts of positive and negative polarity flux, the net flux
and therefore also the net circular polarization is zero. Although the strength
and magnetic energy density of such a tangled field can be arbitrarily high,
it is invisible to the longitudinal Zeeman effect as long as the individual flux
elements are not resolved.
More details can be found in  the work  of  \cite{Johnstone.2010} where the authors  have used simulated stellar surface magnetic maps for  hypothetical stars to discuss   the effects of the loss of information from ZDI maps on the extrapolated coronal fields.

In this paper we focus on the role played by the magnetic flux contained in dark starspots. The missing  flux is simulated by creating spots maps using  information on their location and size  from  stellar brightness maps. We assume  the spots to have  fields {\revised from 500~G to 1~kG} and  a random polarity. These created spots maps are then added to the independently acquired magnetograms.

 By producing many realisations of the spots maps (with random magnetic polarity) we have obtained statistically convincing information on the effect of this missing flux in the stellar coronal extrapolation from the magnetograms acquired using ZDI. 

 

\section[]{Observations and stellar parameters}
In order to characterize  and compare the coronal structure of the stars we consider two stars with magnetic fields that are fundamentally different. The first star is  V374 Pegasi,  a low mass, fully convective M-dwarf, of 0.28 solar masses and 0.28 solar radii. It has a rotation period of 0.446 days and  the angle of the stellar rotation axis to the line of sight is taken to be $70^{\circ}$. The surface magnetogram is derived from detailed spectropolarimetric observations collected with ESPaDOnS on the CFHT in August 2005 by  \citet{Donati.2006b}. This magnetogram shows that the field is predominantly dipolar, with a dipole axis well aligned with the stellar rotation axis. The brightness image shows several dark spots at mid to low latitudes. The second star is AB Doradus (HD 36705) a K0-K2 spectral type star. The mass and radius of this star are very similar to that of the Sun but it has a faster rotation period  (0.514 days). The surface magnetograms are derived from  spectropolarimetric observations collected with the   Anglo-Australian telescope in  December 2004. The stellar rotation axis is inclined by $60^{\circ}$ from  the line of sight. The magnetogram shows a field that is much more complex than that of V374 Peg, with a dipole component having an axis inclined at  $32^{\circ}$  to the rotation axis. The brightness map shows the surface to be heavily spotted with a dark polar cap.

 \begin{table}
\caption{Stellar  parameters  for V374 Peg and AB Dor. } \label{tab1}
\begin{center}
\begin{tabular}{|c|c|c|c|c|c|}\hline
 Star &ST& $\rm{M}_{*}$ ($\rm{M}_{\odot}$) & $\rm{R}_{*}$ ($\rm{R}_{\odot}$) & $\rm{P}_{rot}$ (days) &d (pc)\\
\hline 
V374 Peg&M4 &0.28  &0.28&0.446 & 8.963\\
\hfill\hfill
AB Dor& K1&1&1 &0.514 & 14.94\\ 
\hline 
\end{tabular}
\end{center}
\end{table}

 \begin{table*}
\caption{Three different configurations of spot distribution  added to   V374 Peg's surface, the percentage of the stellar surface area covered by the spots and the spot unsigned flux (normalised to the total unsigned flux in the ZDI map). For each case spots at 500~G and 1~kG were tried. The cases are ordered by decreasing unsigned spot fluxes. }\label{cases_v374peg}
\begin{center}
\begin{tabular}{|c|c|c|c|c|} \hline
case&  brightness  level &spot coverage &normalized spot unsigned flux\\
&&&for spots at  500~G and 1~kG&\\\hline
1& over  0.05&12$\%$&0.03 and 0.13&\\\hline
2& over  0.1&5$\%$&0.01 and 0.06&\\\hline
3& over  0.2&2$\%$&0.005 and 0.02&\\\hline
\end{tabular}
\end{center}
\end{table*}
 \begin{table*}
\caption{Three different configurations of spot distribution  at  AB Dor's surface, the assessed magnetic fields to the added spots, the area covered by the spots and the spot unsigned flux (normalised to the total unsigned flux in the ZDI map). The cases are ordered by decreasing unsigned spot fluxes.}\label{cases_abdor}
\begin{center}
\begin{tabular}{|c|c|c|c|c|c|} \hline
case&  brightness  level &spot field strength (G)&spot coverage &normalized spot unsigned flux&nature of spots\\
&&&&&\\\hline
1&from 0.05 to 0.1&500&5$\%$&2.9&large, high field strength spots\\
&over 0.1&900&30$\%$&&\\\hline
2&from 0.05 to 0.5&500&20$\%$&2.1&large, low field strength spots \\
&over 0.5&900&16$\%$&&\\\hline
3&from 0.4 to 0.7&500&9$\%$&1.43&small spots\\
&over 0.7&900&9$\%$&&\\\hline
\end{tabular}
\end{center}
\end{table*}





\section[]{Surface field geometry\label{spot-maps}}

\begin{figure*}
\resizebox{18.5cm}{!}{
\includegraphics[angle=90.]{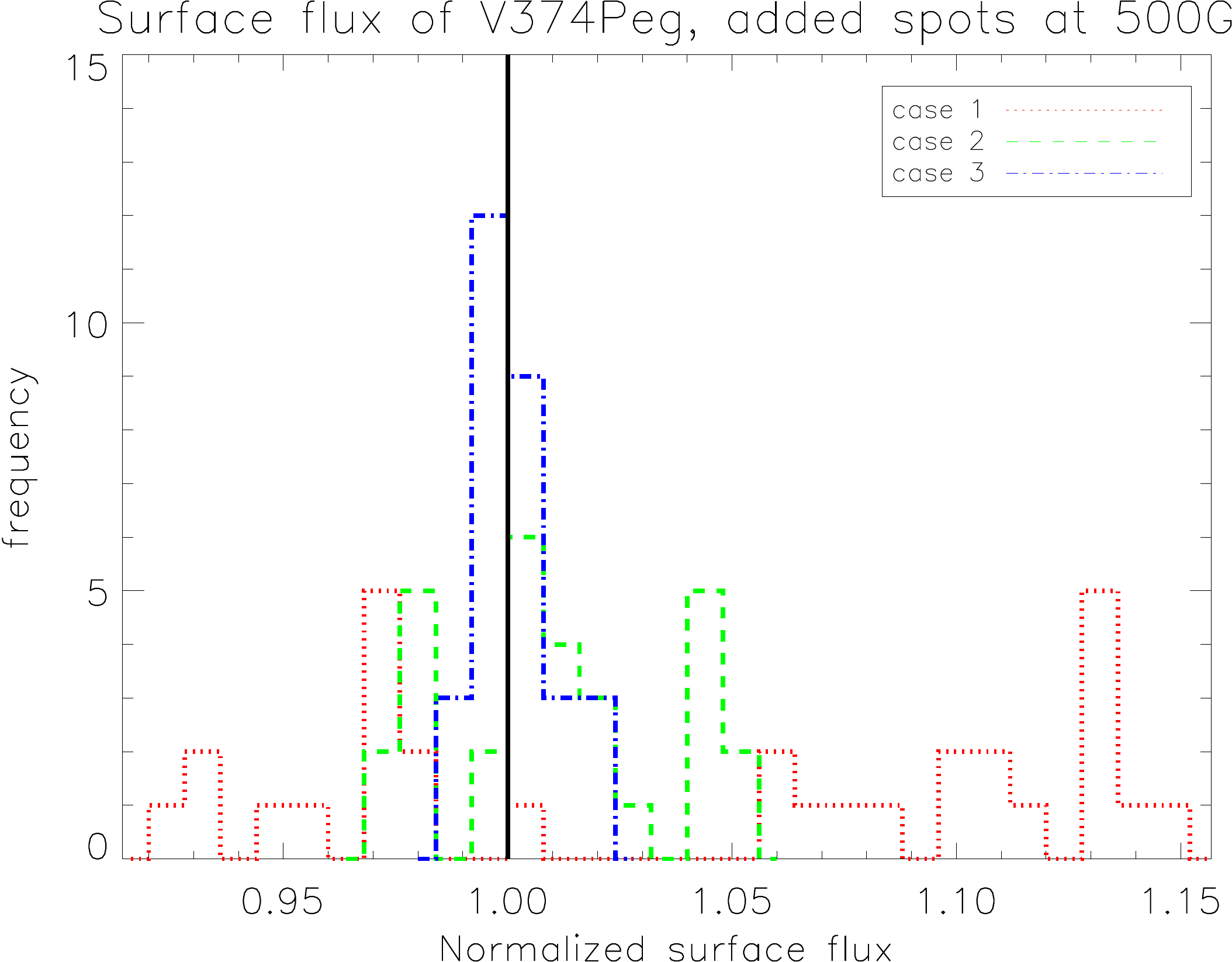} 
\includegraphics[angle=90.]{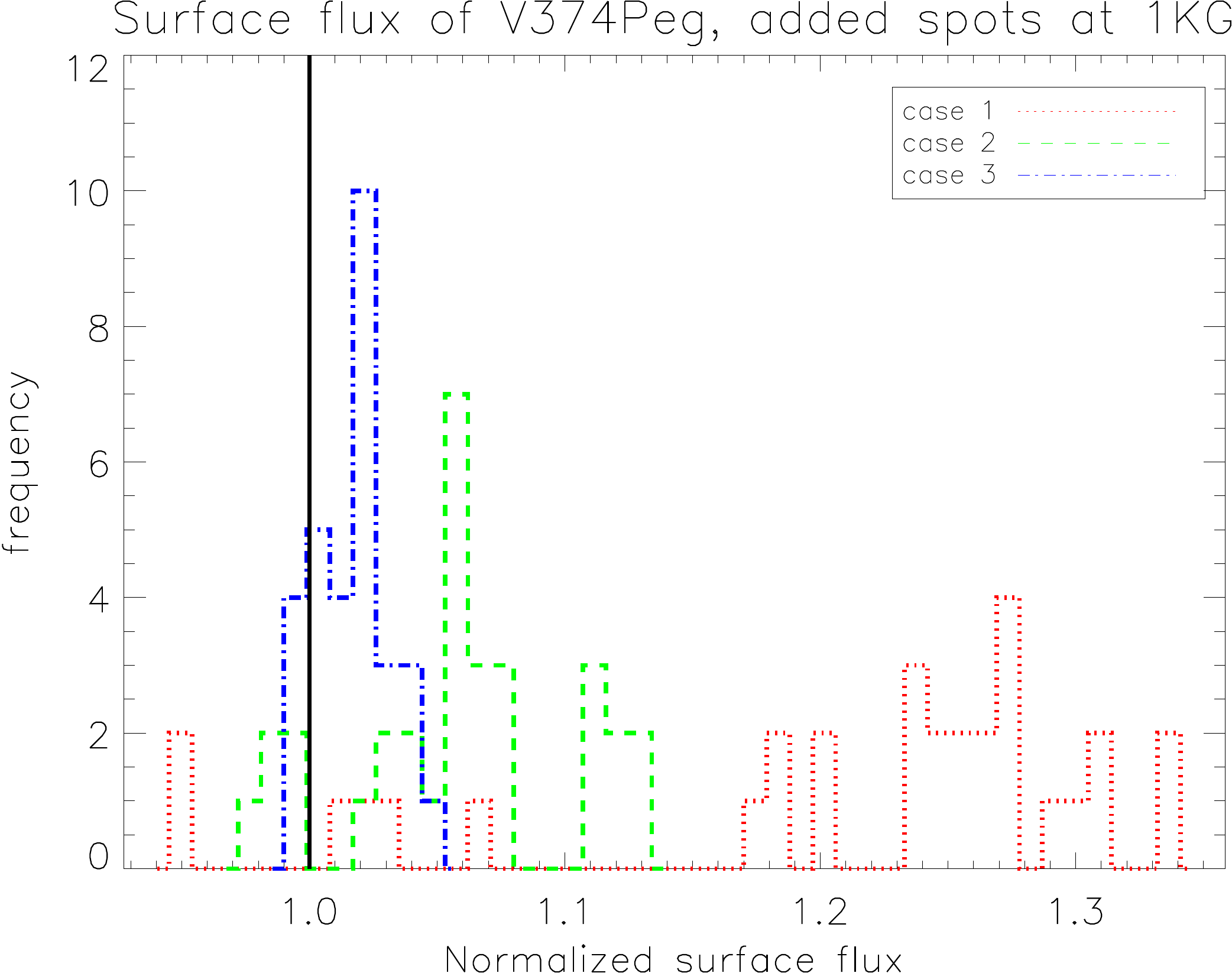} 
\includegraphics[angle=90.]{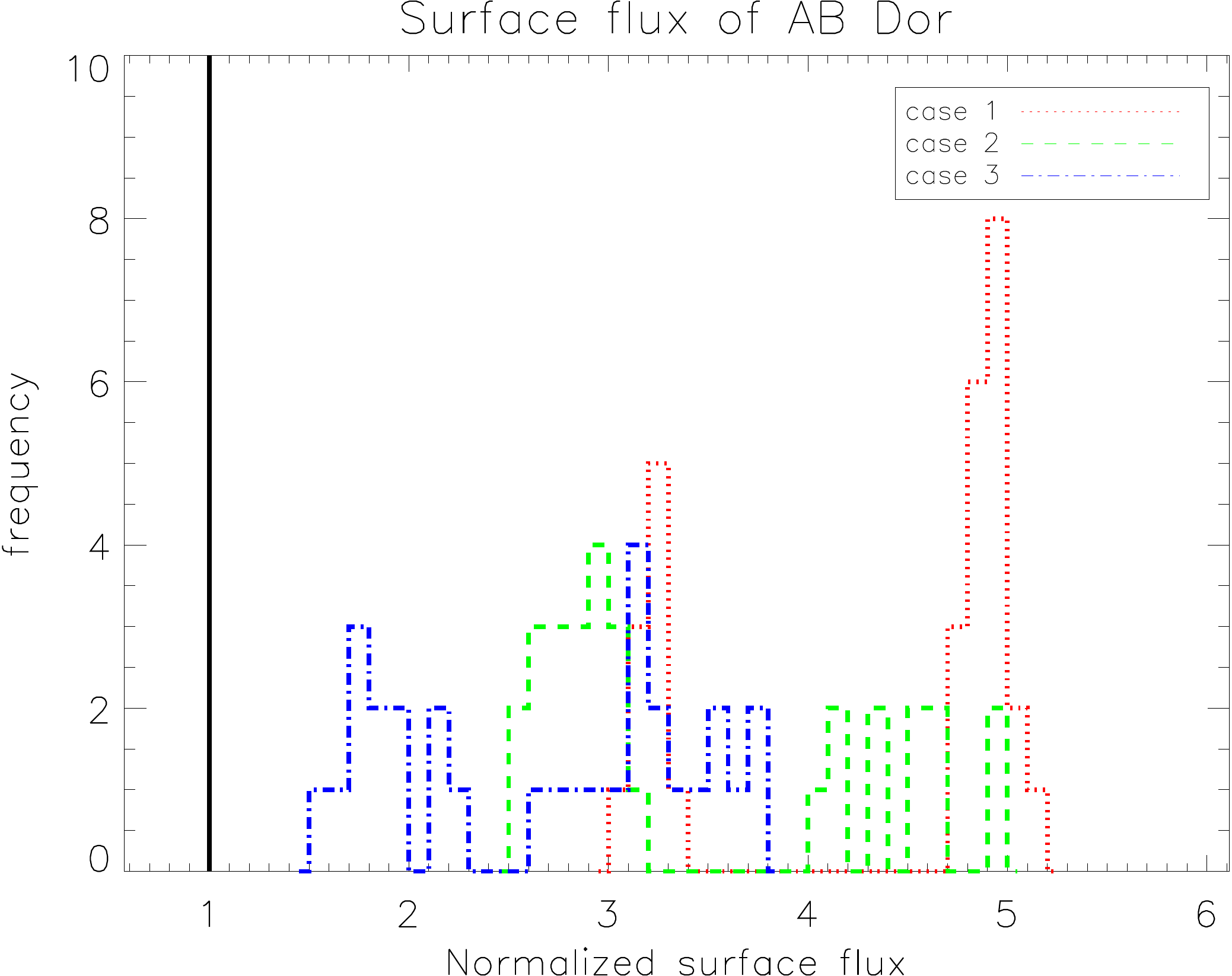}}
\caption{Normalized total unsigned  flux after adding spots maps. The first 2 plots correspond to V374 Peg. Spots have been added with field strengths of 500~G (left) and 1000~G (middle). The different histograms  correspond to  the different cases in Tables \ref{cases_v374peg} and \ref{cases_abdor} and the straight vertical black line corresponds to the value of the ZDI map. The plot on the right corresponds to the 3 different realizations of spots for AB Dor. }
\label{total_flux}
\end{figure*}

We do not know {\em a priori} the magnitude or polarity of  the magnetic flux in the dark spotted regions. Guided by the solar analogy, however, we can consider several illustrative examples. We choose to vary both the spot area (characterised by the level of the brightness contour defining the spot) and the field strength within that spot. By taking a range of spot areas and field strengths we can explore the nature of the effect that this extra flux might have.

 \subsection{Characterising the ``spot flux''}

For V374 Peg we consider two different field strengths: 1000~G  and a much weaker spot field of 500~G. Spots are defined by brightness contours set at a range of 
values (0.05, 0.1 and 0.2). The resulting magnetograms due to the spot field can be seen in  Fig.~\ref{0805spots}. 
The effect of adding  spot-maps to the ZDI original map  is obvious in the sample modified magnetogram (Fig.~\ref{0805disp}). The dipolar shape gives place to a less organized  field, with several spot-shape regions in the upper hemisphere. The detailed appearance depends of course on the particular realisation of the spot-map, on the scale of the created spots and the allocated field strength and polarity. 
   
In the case of AB Dor, similar steps were taken to create spots maps and modified magnetograms for AB Dor. In this case, however, defining the spots is more complicated because of the larger spot coverage  (see the corresponding brightness map of AB Dor  in Fig.~\ref{ABdor}).  We tried three different cases to describe some possible spot distributions and their magnetic field strengths. The first case describes high field strength spots that extend over a large fraction of the surface area. This is achieved by allocating a field strength of 500~G to brightness levels between 0.05 and 0.1 and a field strength of 900~G to brightness levels over 0.1.  The second  case represents low field strength spots that extend over a large fraction of the surface area. This is achieved by allocating a field strength of 500~G to brightness levels between 0.05 and 0.5 and a field strength of 900~G to brightness levels over 0.5.  The third case describes small spots - here spots with a field strength of 500~G are allocated to brightness levels between 0.4 and 0.7 while a field strength of 900~G is allocated to brightness levels over 0.7. Examples of spot-maps deduced by the brightness map considering the three different cases detailed above can be seen in Figures~\ref{0805spots}  and \ref{ABDor_spots}.

These three cases introduce different spot scales to the ZDI map as well as different polarities.

\subsection{ Total magnetic flux added in spots}

The degree to which the magnetic flux from the spots can influence the geometry of the coronal magnetic field depends both on the magnitude of the spot flux that is added (relative to the magnetic flux from the ZDI map) and also the polarity that is added, since,  if this is opposite to the magnetic polarity of the surrounding region, the net effect might be cancellation rather than addition.  We therefore calculate the total unsigned flux through the stellar surface for each realisation of the spot field. Fig.~\ref{total_flux} represents a statistical distribution of the total flux of V374 Peg and AB Dor after adding the spots to the original ZDI map. These values are normalized by the total flux of the corresponding naked ZDI map (represented by the straight line equal to  the value one). In the case of V374 Peg, the flux in the spots is a small fraction of the flux in the ZDI map. When the spot flux is extremely small (as in case 3 in the left hand panel of Fig.~\ref{total_flux}) its effect is simply to cause a small perturbation and the histogram of possible values of the unsigned flux is centred around 1. As the magnitude of the spot flux is increased, however, this distribution becomes skewed towards larger values.  For the largest spot flux (case 1) the total flux values vary between 0.9 and 1.13 (spots at 500~G) or 0.85 and 1.27 (spots at 1000~G). In the case of AB Dor, however, the magnitude of the magnetic flux in the spots is  much greater than that in the ZDI maps.  In this case,  the addition of spot field to the ZDI map of AB Dor can increase the total flux significantly. It is notable that for AB Dor the distribution appears bimodal. This is likely to be due to the distribution of spot sizes on AB Dor, which is dominated by the large polar spot. The two peaks of the histogram correspond to the two possible polarities for this polar spot. 

\section[]{Magnetic field extrapolation}

\begin{figure*}
\resizebox{18.5cm}{!}{
\includegraphics[angle=90.]{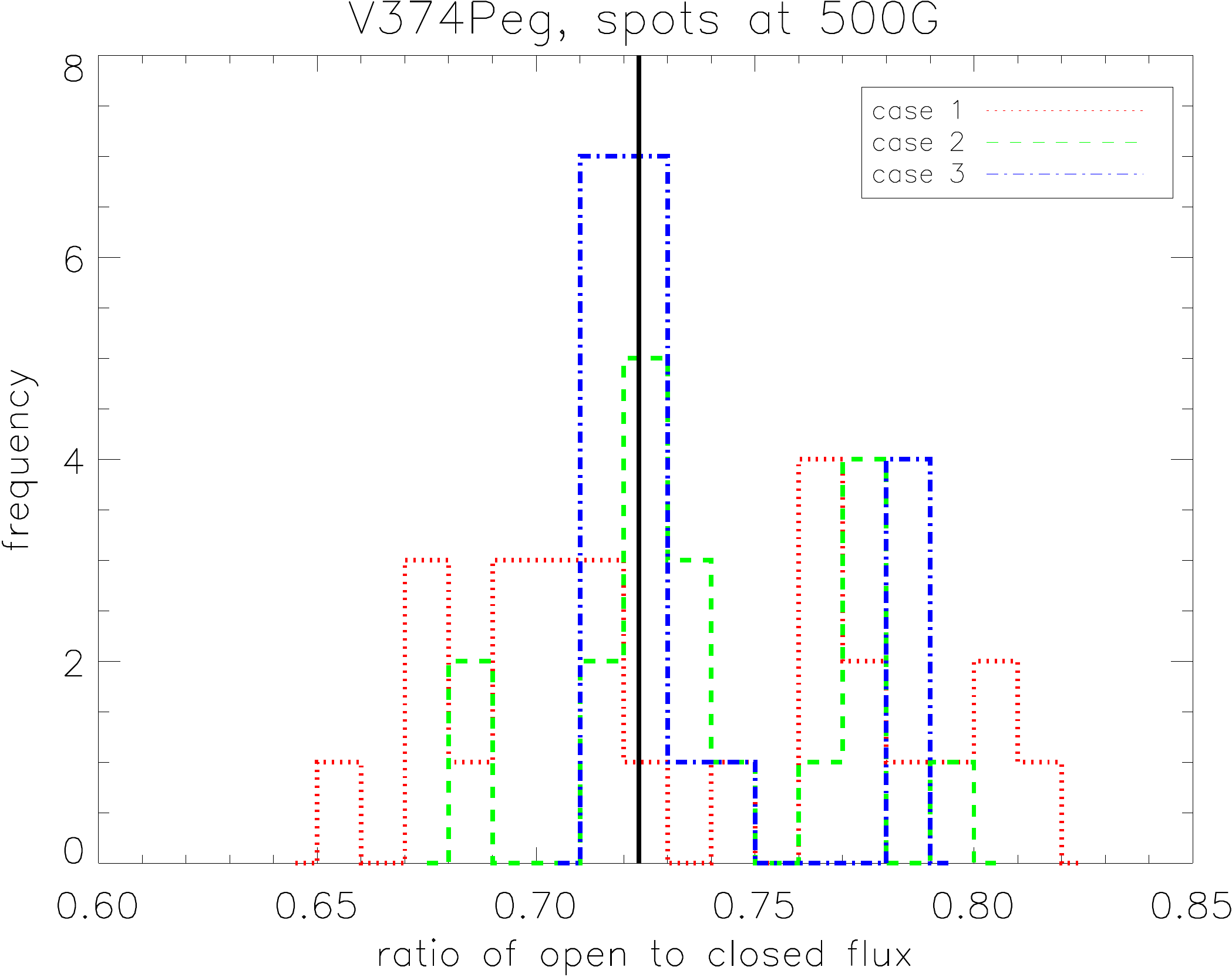} 
\includegraphics[angle=90.]{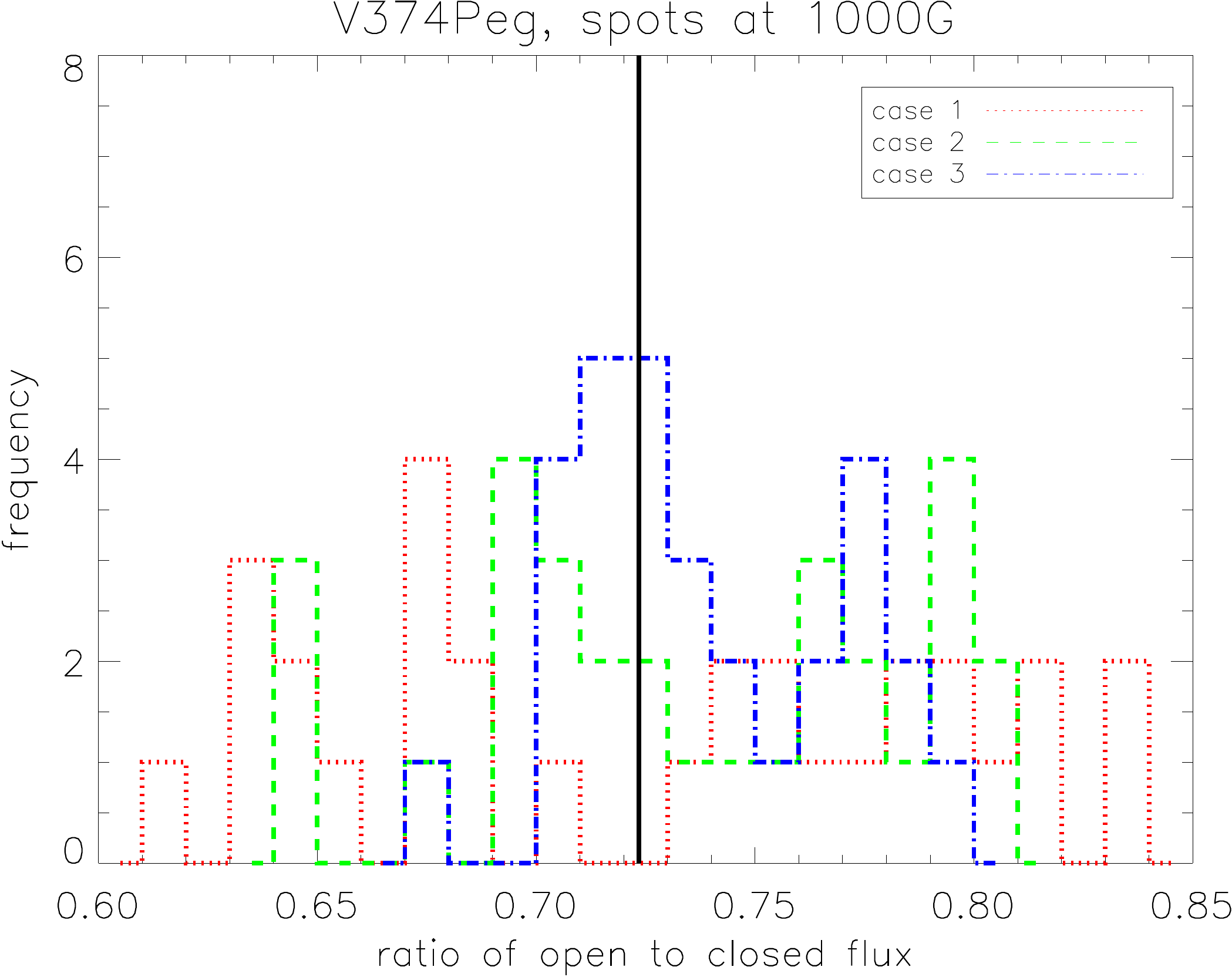} 
\includegraphics[angle=90.]{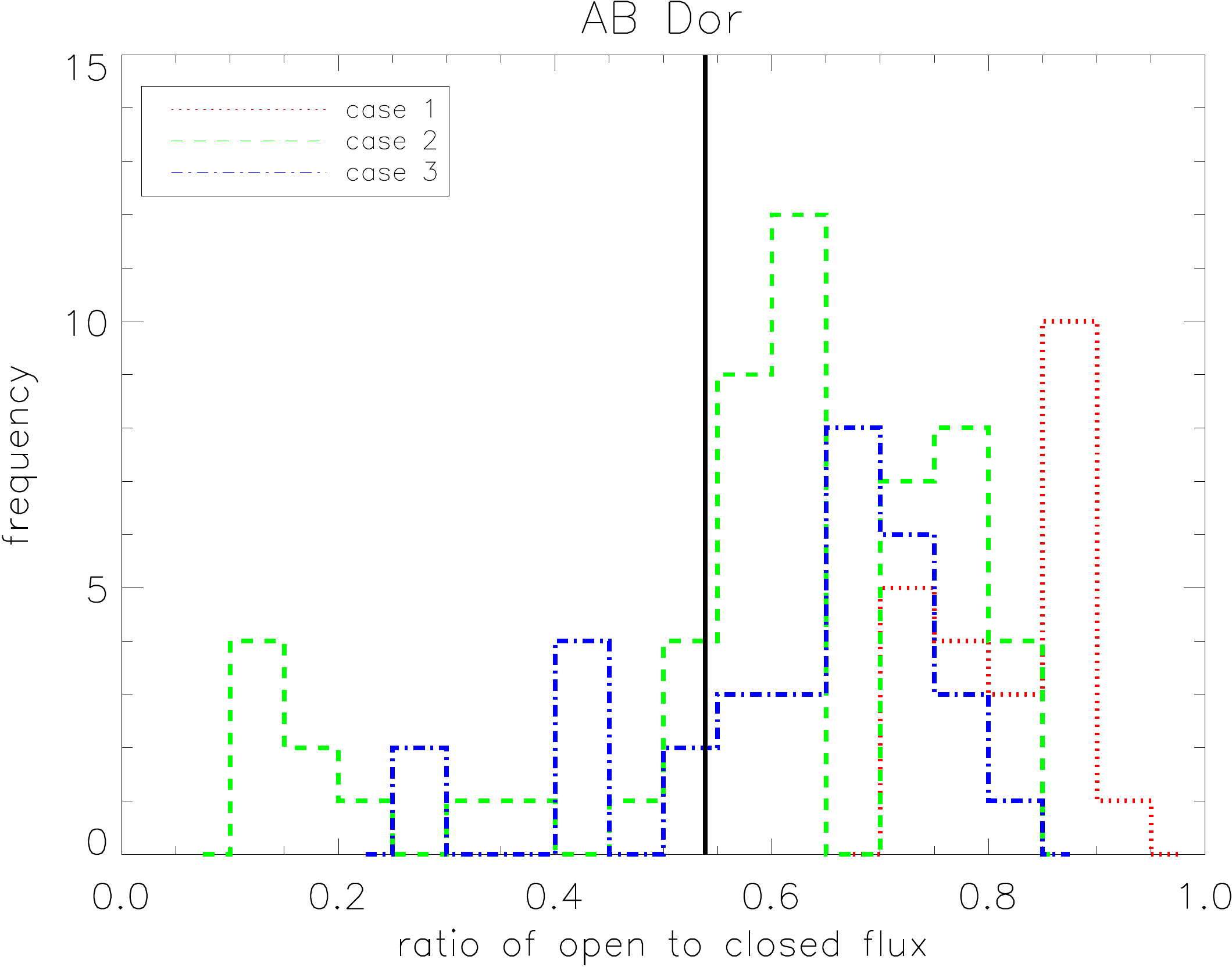}}
\caption{Ratio of coronal open and closed {\revised flux}. The first 2 plots correspond to V374 Peg. On the left panel, the field strength in the spots was 500~G, while in the middle panel, it was1000~G. The different colours  correspond to  the different cases and the straight vertical  line correspond to the value of the ZDI map. The plot on the right corresponds to the 3 different realizations of spots for AB Dor. }
\label{ocf}
\end{figure*}

From the derived maps of the photospheric fields of the stars we extrapolate their three-dimensional coronal field structure  using the potential field source surface extrapolation model.  The magnetic  field is described as a sum of multipole components represented in  spherical polar coordinates ($r$, $\theta$, $\phi$) with 0 the origin of the star \citep{Willis.1987}. The field is current-free.

As we have assumed that the field is potential, then $\nabla \times B =0 $. This condition can be satisfied by writing the field in terms of a scalar flux function $\psi$ such that 
\begin{equation}
B = -\nabla \psi .
\end{equation}\label{B}
The field must also satisfy Maxwell's equation, $\nabla.B = 0$, therefore $\psi$  must satisfy Laplace's equation \citep{Chandrasekhar.1961}, 
\begin{equation}
\nabla^{2} \psi = 0 .
\end{equation}
The solution of this equation in spherical coordinates is written as,

\begin{equation}
\psi=\sum_{\ell=1}^{N}\sum_{m=-\ell}^{\ell} \left[a_{\ell m}r^{\ell}+b_{\ell m}r^{-(\ell+1)}\right]\,\,P_{\ell m}(\theta) e^{im\phi}
\end{equation}
where $P_{\ell m}$ are the associated Legendre functions of  harmonic degree $\ell$ and  azimuthal mode $m$ (note, the $\ell$ = 0 component
is not considered as it corresponds to a purely radial  field component which
does not exist in the absence of magnetic monopoles). This can be solved to
give the three components of the  field in terms of a set  of constants $a_{\ell m}$ and
$b_{\ell m}$ using

\begin{eqnarray}
B_{r}&=&-\sum_{\ell =1}^{N}\sum_{m=-\ell }^{\ell } \left[la_{\ell m}r^{\ell -1}-(\ell +1)b_{\ell m}r^{-(\ell +2)}\right]\,\,P_{\ell m}(\theta) e^{im\phi}\label{blm1}\\
B_{\theta}&=&-\sum_{\ell =1}^{N}\sum_{m=-\ell }^{\ell } \left[a_{\ell m}r^{\ell -1}+b_{\ell m}r^{-(\ell +2)}\right]\frac{d}{d\theta}P_{\ell m}(\theta) e^{im\phi}\label{blm2}\\
B_{\phi}&=&-\sum_{\ell =1}^{N}\sum_{m=-\ell }^{\ell } \left[a_{\ell m}r^{\ell -1}+b_{\ell m}r^{-(\ell +2)}\right]\frac{P_{\ell m}(\theta) }{sin\theta}ime^{im\phi}.
\label{blm3}
\end{eqnarray}

\begin{figure*}
\resizebox{18.5cm}{!}{
\includegraphics[angle=90.]{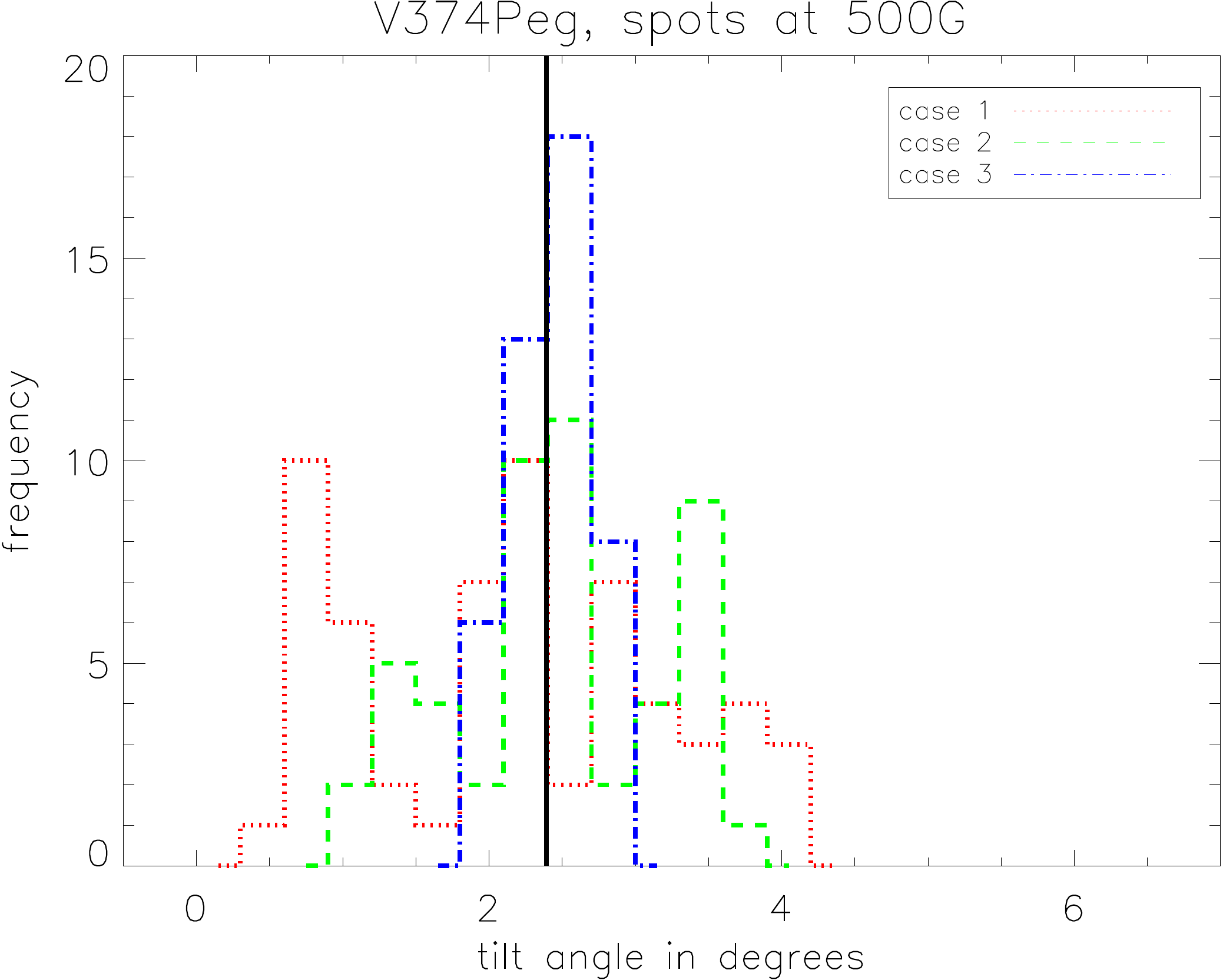} 
\includegraphics[angle=90.]{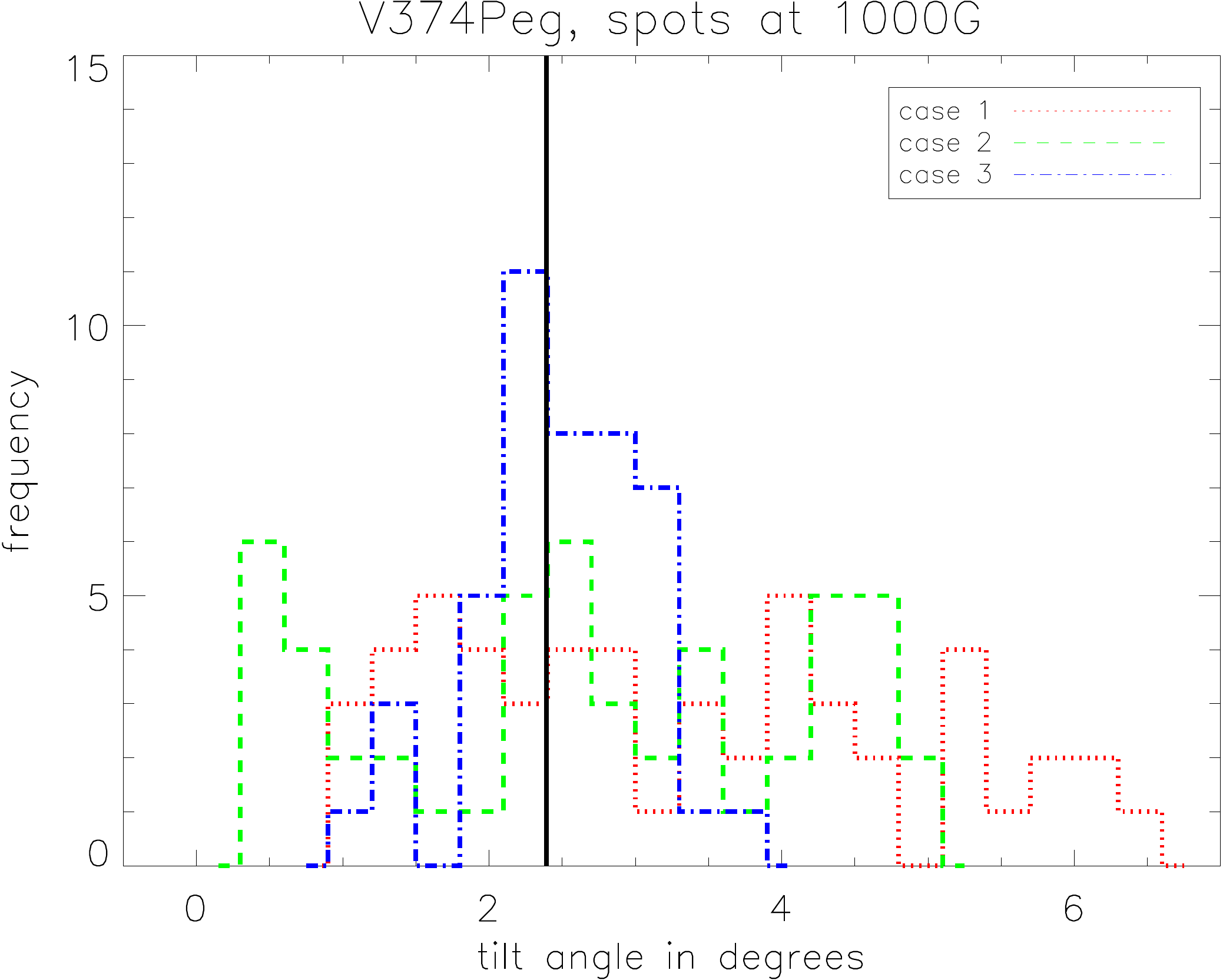} 
\includegraphics[angle=90.]{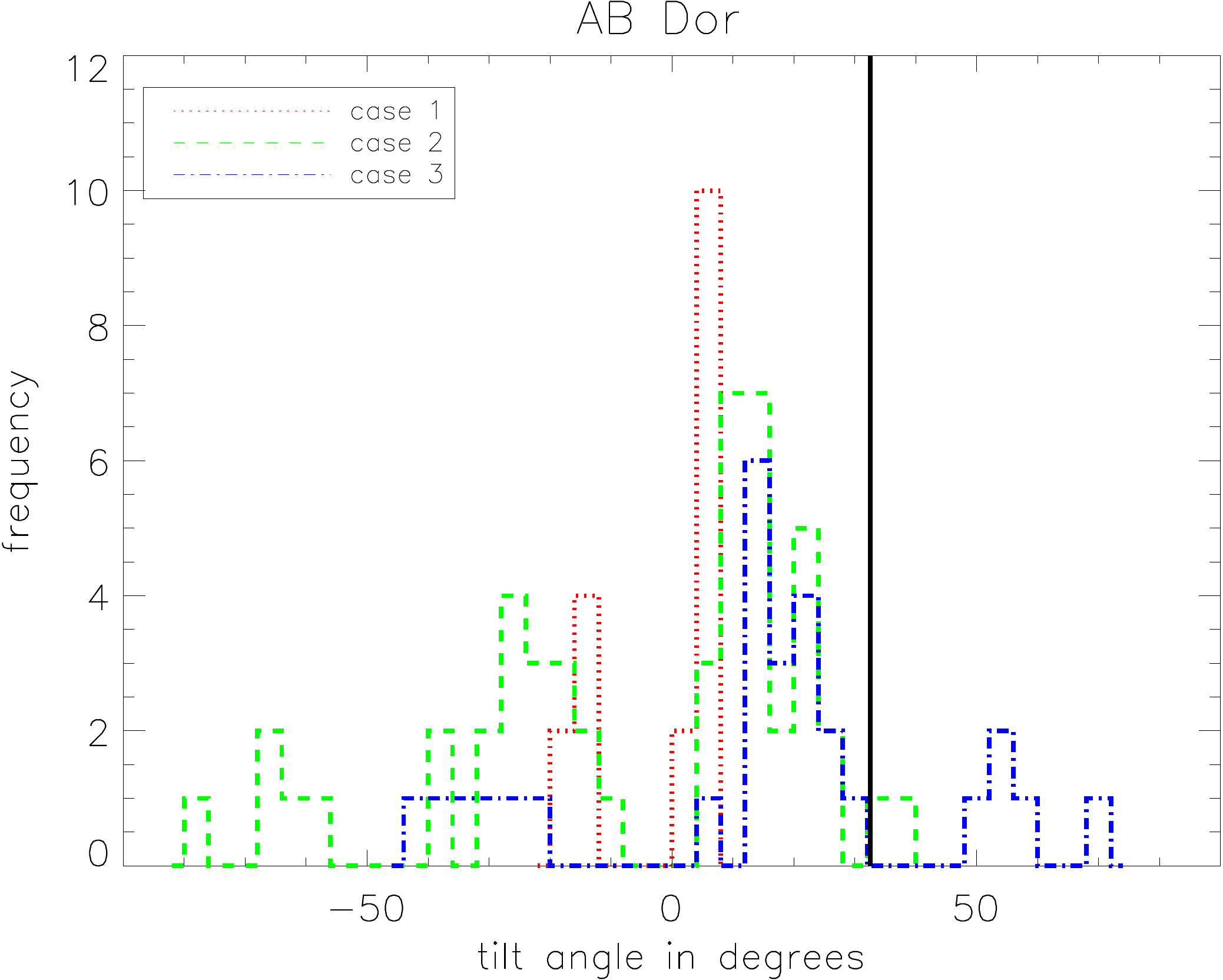}}
\caption{Tilt angle of the dipolar component of the field. The first 2 plots correspond to V374 Peg and show the effect of the addition of spots at 500~G (left) and 1000~G (middle).The different histograms  correspond to  the different cases in Table \ref{cases_v374peg} and the straight vertical black line corresponds to the value from the ZDI map. The plot on the right corresponds to the 3 different realizations of spots for AB Dor. }
\label{tilt_angle}
\end{figure*}

Two boundary conditions are needed in order to determine  the spherical harmonic coefficients $a_{\ell m}$ and $b_{\ell m}$. The first boundary condition is the strength of the radial  component of the field at the stellar surface that is derived from the Zeeman-Doppler maps of the star. The second condition is that  at some height above the stellar surface $R_{s}$, known as the \textit{source surface},  where {\revised the coronal magnetic field  opens} to form the stellar wind, the field becomes radial and hence $B_{\theta}(R_{s})=B_{\phi}(R_{s})=0$ \citep{Altschuler.1969}. Thus it is possible to determine the magnetic field components $B_{r}$, $B_{\theta}$ and $B_{\phi}$ and therefore a field vector at any point within the corona.

The magnetic field components are  given as the sum over several terms, each corresponding to individual values of $\ell$ and
$m$. Each term corresponds to a different associated Legendre function and represents a
single simple  field configuration. Thus if we consider axisymmetric ($m$=0) fields,  a purely radial  field has $\ell$ = 0, a dipole
has $\ell$ = 1 and a quadrupole has $\ell$ = 2. 

As can be seen in the equations \ref{blm1}, \ref{blm2} and  \ref{blm3}, complex potential  fields are described as the
superposition of many of the individual modes, with the power, and thus the 
influence, of each mode given by the $a_{\ell m}$ and
$b_{\ell m}$ coefficients. Thus, in the
potential field source surface  model, a global magnetic  field is completely described by its corresponding
set of  $a_{\ell m}$ and $b_{\ell m}$ values.  More complex  fields   are described
by higher order modes.
It is important to note also  that the strength of higher order modes fall off   with
increasing distance from the surface of a star faster than lower order. Thus, more
complex fields  (i.e. higher $\ell$ value)  are only able to affect coronal plasmas
 to smaller radii than less complex  fields. This also means that
different components of a  field may dominate at different distances, for instance
with a quadrupolar  field dominating close to the stellar surface and a dipolar
 field dominating at large radii.
 
The field extrapolation was performed  using a code originally developed by \citet{vanBallegooijen.1998}. The source surface radius is taken to be 3.4 solar radii  and the extrapolation in spherical harmonics extends to $\ell$=31.

\section{Coronal field geometry}

The influence of the magnetic flux from the spots is different for the two stars we are considering. This is a combination of the difference in the ZDI field of each star and the magnitude and spatial distribution of the flux in the spots. From the ZDI map alone, it is clear that V374 Peg has a strong, dipolar field with {\revised a strong  north-south flux} - indeed it  {\revised has much more magnetic flux that crosses the equator} than AB Dor. In addition, the spots on V374 Peg, which contribute a small fraction to the total magnetic flux,  are at low latitudes and distributed fairly evenly in longitude. This alone suggests that the spots might have only a small effect on the geometry of the coronal magnetic field. By comparison, the ZDI map of AB Dor shows  to have a much more complex field, with {\revised more  East-West flux} than V374 Peg. In addition, it has a very pronounced polar spot and the contribution to the total flux from the spots is much greater. 
We choose to study two simple measures of the geometry of the coronal magnetic fields of these stars. The first is the ratio of open to closed flux. This is important as the open field lines are those   that carry   the stellar wind and are  therefore able to contribute to the angular momentum loss that spins the star down as it evolves. The high rotation rates of many low mass stars such as V374 Peg may be related to the efficiency with which their winds can carry away angular momentum and hence the fraction of the stellar magnetic flux that can contribute to such a wind is a crucial parameter. The  second measure of the field geometry is the tilt angle of the large-scale, dipolar component of the field. In the case of the Sun, the reversal of the dipole axis over the course of a magnetic cycle is a clear indicator of the nature of the dynamo-generated field. The drift of the dipole axis is driven by the emergence and evolution of new, mainly East-West flux in the form of the bipolar regions that form spot pairs. The addition of this flux to the mainly North-South field present at the start of a new cycle eventually tilts the axis of the large-scale field into the equatorial plane until it is completely reversed. In the case of the Sun, both the dipole tilt and the fraction of open flux vary through the magnetic cycle. It is therefore important to know, in the case of other stars, what contribution the spot field might make to their derived values.


\subsection{Ratio of open to closed field}

To have a first indication of the global coronal structure we calculated  the ratio of  open to closed field in the corona. In the case of V374 Peg, the addition of the magnetic flux in the small scale, low latitude spots can clearly either increase or decrease the fraction of open flux. The range of variation increases with the flux of the added spots (see Fig.~\ref{ocf}a).   

 This effect is much more obvious in the case of AB Dor.  When adding  large scale spots with strong field, the open flux  can increase by as much as $30\%$. Although in some realisations of the coronal field, the fraction of open flux is decreased when the spots are added, the effect is more often to increase the open field. This may come from the presence of the large scale polar spot,  visible in the brightness map, which largely contributes to the open field (see the ZDI map figure~\ref{ABdor}). In the most extreme cases, the increase in the open flux is so large that it would impact significantly on the angular momentum loss. The ``opening up'' of the polar cap magnetic flux  would also ensure that the pole was dark in X-ray emission, thus potentially influencing the rotational modulation of the X-ray emission.

\begin{figure*}
\resizebox{18.5cm}{!}{
\includegraphics[angle=90.]{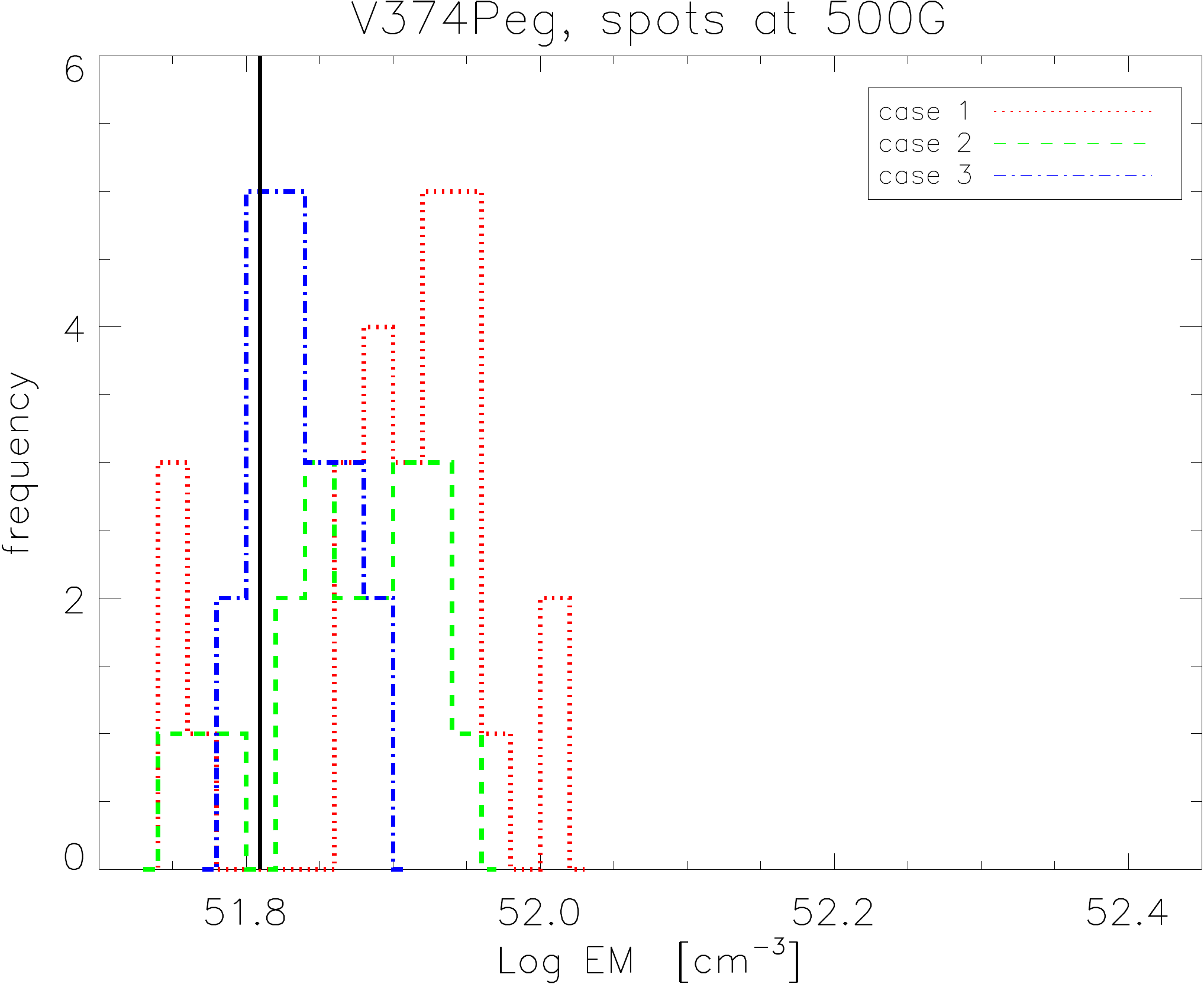}
\includegraphics[angle=90.]{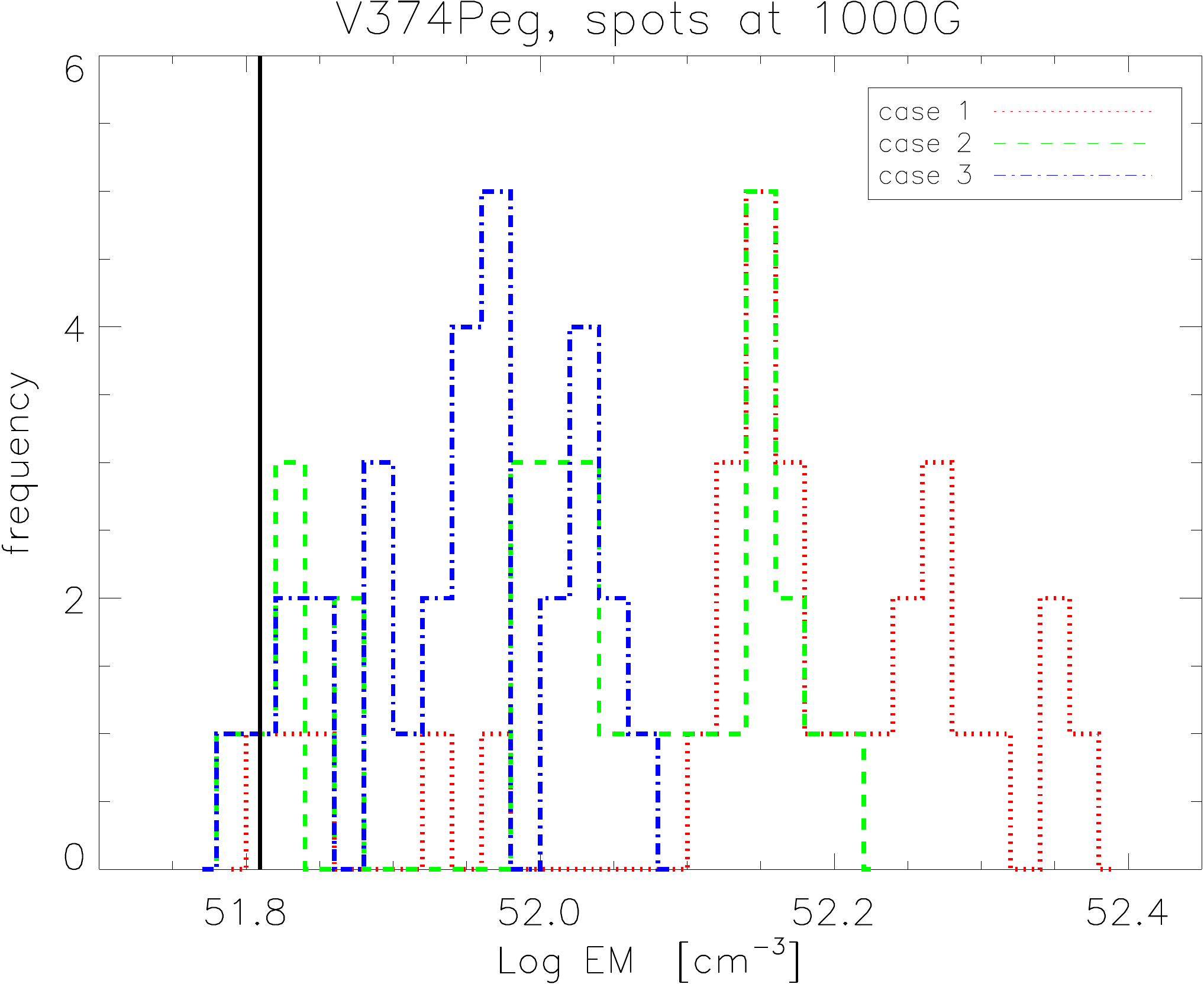}
\includegraphics[angle=90.]{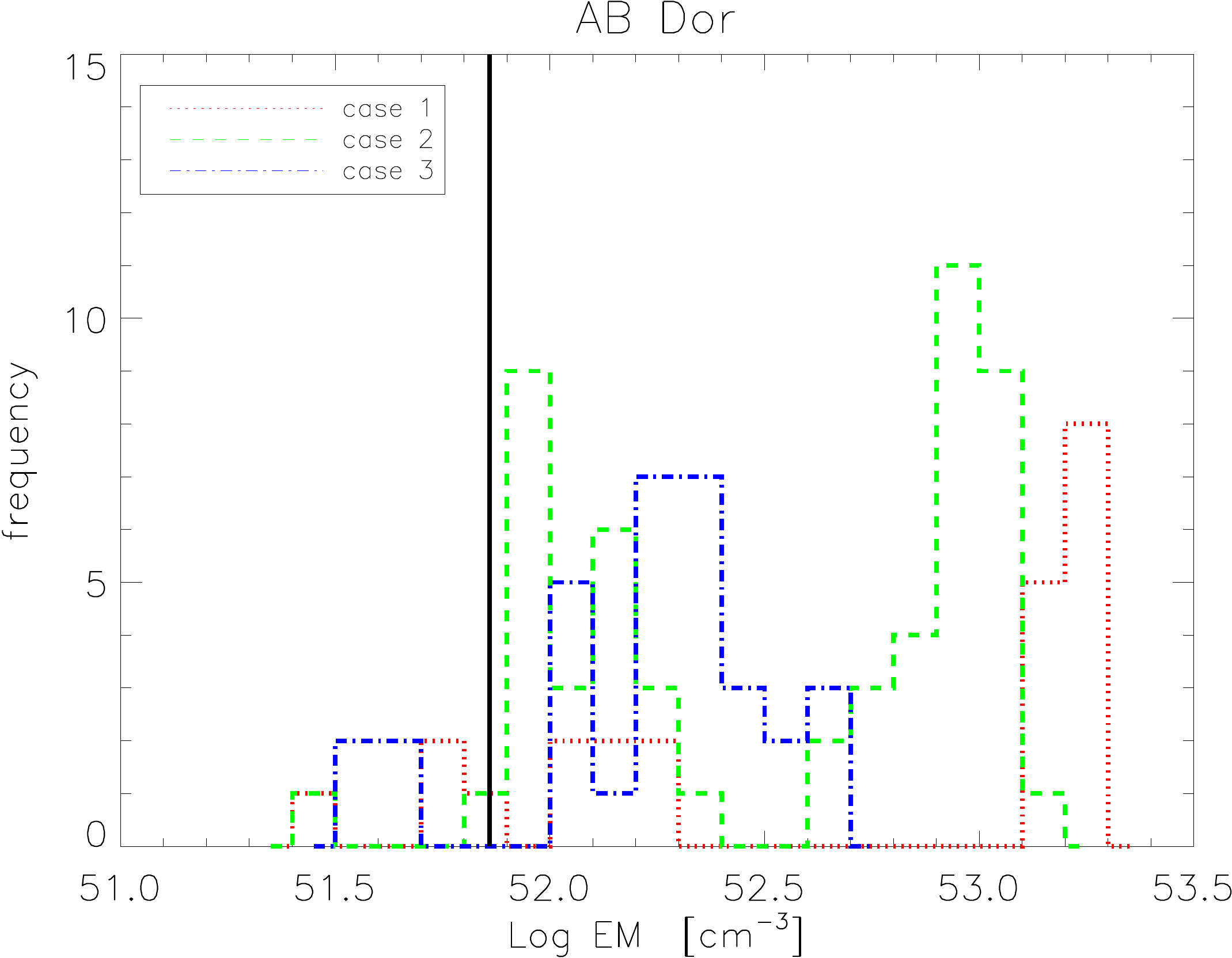}
}
\caption{X-ray emission measure  for both stars. The first 2 plots correspond to V374 Peg. on the left added spots at 500~G in the middle at 1000~G, the dots are the values for each realisation and the straight horizontal line corresponds to the value of the ZDI map. The plot on the right corresponds to the 3 different realizations of spots for AB Dor.}
\label{EM}
\end{figure*}

\subsection{ Tilt angle of the dipole component}

The second measure of the field geometry is the angle between the dipole component of the magnetic field, which defines the large-scale field, and the stellar rotation axis. AB Dor has a distorted dipole component which is tilted by  $32.5^{\circ}$  from its rotation axis. Whereas 
for V374 Peg it is just  tilted by $2.4^{\circ}$. These values are changed when adding spot maps to the ZDI maps (see Fig.~\ref{tilt_angle}).

In the case of V374 Peg the  tilt angle changes randomly on both sides of  the original value, shifting further from the rotation axis when larger spot fluxes are added. This is in agreement with the decrease of {\revised the magnetic flux} crossing the equator  which shows a weakening of the axisymmetric dipolar component of the field and  the emergence of magnetic structures closer to the equator.

On the contrary,  the tilt angle of the dipolar component of  AB Dor's field moves closer to the rotation axis, especially when the spot fluxes are large(first case; red dashed  histogram in Fig.~\ref{tilt_angle}c). The polarity of the added spots influences also the tilt angle of the dipole component shifting it   from values  between $-80^{\circ}$ to $60^{\circ}$ comparing to  the rotation axis, with mean tilt angle values of $-4.3^{\circ}$, $10.5^{\circ}$ and $-2^{\circ}$ respectively  for the three cases.  From AB Dor's brightness map comes  the addition of a large scale spot at one pole of the star which induces a strong axisymmetric dipolar field. This magnetic region  is absent from the ZDI map and changes the coronal structure of the star.  

\section{Structure of the coronal plasma} 

Clearly, the addition of the spot field can change the geometry of the coronal magnetic field, altering the locations of closed field (capable of containing hot plasma) and open field (capable of carrying a wind). This will in turn affect the density structure of the coronal plasma and hence both the magnitude and rotational modulation of the X-ray emission measure. 

\subsection{Coronal gas pressure and  density}

We calculate the pressure  of the corona assuming  it to be isothermal and in hydrostatic equilibrium  where $\vec{\nabla} P= \rho \, \vec{g}$. The pressure at any point along a magnetic field line is then given by


\begin{equation}
P(r,\theta,\phi)= P_{0}\exp\left(\frac{m}{ k_{B}\, T} \int g_{s}dS\right),
\end{equation}
where $m$ is the mean mass of the particles in the plasma, $T$ is the temperature of the corona set to be 10M~K and  $k_{B}$ Boltzmann's  constant.
$g_{s}$ is the component of the effective gravitational acceleration along the  field
line given by

\begin{equation}
g_{s} =  \frac{\vec{g} . \vec{B}}{|B|}.
\end{equation}
The vector $\vec{g}$ does not just represent the action of gravity on the coronal
plasma, but also takes into account the centrifugal force due to the rotation of
the star. Other accelerating mechanisms, such as radiation pressure, are ignored
here. Thus, g is given by

\begin{equation}
g(r,\theta, \phi) = ( -GM/r^{2} + \omega^{2}\,r\, \sin^{2}\theta,  \omega^{2}\,r\, \sin\theta \,\cos \theta, 0).
\end{equation}

At the footpoints we scale the plasma pressure to the magnetic pressure such that   the pressure at the surface of the star is proportional to the square of the surface magnetic field strength:
 
\begin{equation}
P_{0} \propto B_{0}^{2} .
\end{equation}
The ratio between the pressure at the surface of the star and  the surface magnetic pressure  
 is denoted  by K. \citet{Jardine.2006} provides a detailed explanation of how $P_{0}$, the coronal base pressure, can be scaled to the magnetic pressure at a field line foot point. In this work this value is taken to be $3 \times10^{-6}$.
The plasma pressure within any volume element of the corona is set to zero if the field  line is open, or  if at any point along the field line the plasma pressure is greater then the magnetic pressure, which means  $\beta>1$ (where $\beta =P/P_{mag}$ with $ P_{mag}=\frac{B^{2}}{2\mu_{0}}$, $\mu_{0}$  being the vacuum permeability or  the magnetic constant). This mimics the effect of a high gas pressure forcing  closed magnetic field lines to open up.

From the pressure we calculate the density in the corona assuming the plasma to be an ideal gas 

\begin{equation}
n_{e}=P/ \, k_{B}\, T.
\end{equation}
The coronal mean density is then evaluated for each star using the new spot-magnetograms. 

%

For V374 Peg, the values of the mean densities typically increase after adding spots to the ZDI original map by an amount determined by the magnitude of the magnetic flux added (since the base pressure and density both scale with the square of the magnetic field strength).
On the contrary, in the case of AB Dor, the coronal mean density can either increase or decrease the original value obtained by the ZDI naked map. This may be due to {\revised the increase of open flux} because of the addition of  magnetically strong, high latitude spots. The added  magnetic field opens  a greater fraction of coronal flux,  allowing more plasma to escape as a stellar wind, and hence decreasing the value of the coronal mean density.

\subsection{ X-ray emission measure} 



The optically thin X-ray Emission Measure is determined  by integrating the squared density along lines of sight through the corona.
 
 \begin{equation}
L_{X} \sim EM=\int_V n_{e}^{2}\, dV.
\end{equation}
Since the emission  depends on the square of the density it must be greatest in the dense regions. The pressure is always higher near the star's surface, so is the density of the plasma and consequently  the main  emission measure comes from small structures close to the corona.

The variation of the X-ray emission measure follows closely the variation of the coronal density. The X-ray emission of V374 Peg calculated using the ZDI naked map is $6.5\times10^{51}$~cm$^{-3}$,  the values vary between $(5.5 - 10)\times10^{51}$~cm$^{-3}$ when the added spots are at  500~G and between $(6 - 20)\times10^{51}$~cm$^{-3}$ when the added spots are at  1~kG. In the case of AB Dor the X-ray emission measure varies between $(0.3 - 15)\times10^{52}$~cm$^{-3}$ while the value extracted form the original ZDI map is $0.7\times10^{52}$~cm$^{-3}$.

\begin{figure}
\resizebox{7.5cm}{!}{
\includegraphics{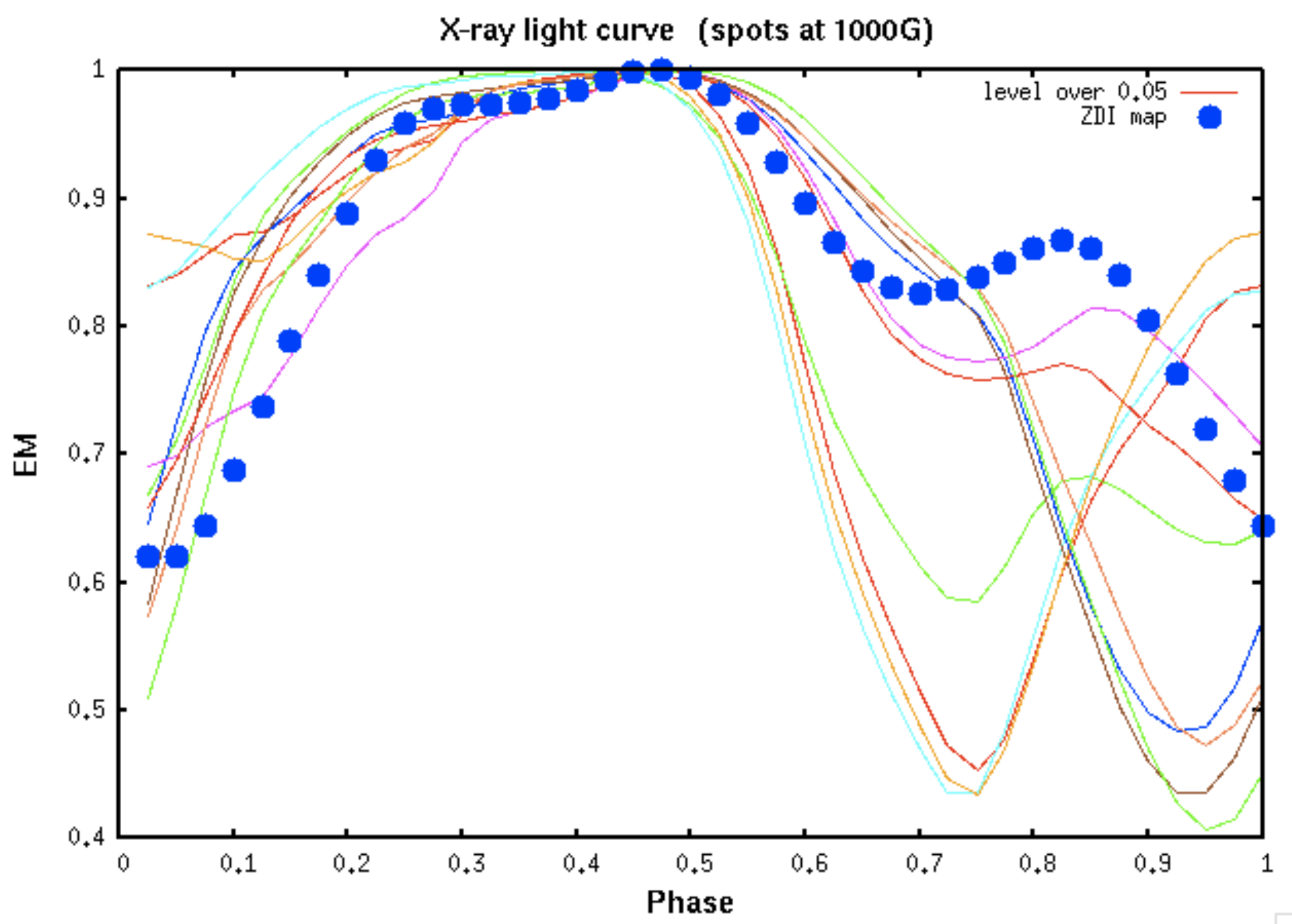}}
\caption{X-ray rotational modulation corresponding to V374 Peg. The dots represent the modulation calculated from the  ZDI map. The lines represent the  X-ray rotational modulation  using  the created maps (ZDI + spot maps) with spots defined by a brightness level above 0.05 with a magnetic intensity of 1000~G.}
\label{005rotmod}
\end{figure}

\subsection{ Filling factor in  the corona} 

Clearly, considering the magnetic field  that may be present in the spots increases the predicted X-ray emission measure, by almost one order of magnitude for V374 Peg and two orders of magnitude for AB Dor. This might at first sight be surprising, given that as shown in Fig.~\ref{ocf}, the addition of the spot field typically increases the fraction of the flux that is open, and hence not contributing to the X-ray emission measure.  The reason can be seen by calculating the filling factor of emitting gas, defined by

\begin{equation}
ff=\frac{\int n_{e}^{2}\, dV}{\overline{n}_{e}^{2}\,(4/3 \, \pi (R_{s}^{3}-R_{*}^{3})) }
\end{equation}
which represents  the calculated emission measure as a fraction of the emission measure of a sphere extending to the source surface $R_{s}$ and uniformly filled with plasma at the mean coronal density. For an isothermal model, this filling factor simply measures the fraction of the coronal volume that is emitting.

For V374 Peg the filling factor typically decreases when spots are added to the ZDI maps - a result that is even more evident when the added spots have higher field strength. This suggests that the emission is originating from smaller structures within the corona when the spot field is considered. This is because the typical lengthscales of {\revised the magnetic field}  associated with the spots are smaller than the very large-scale field of the overall dipole described by the ZDI map. By comparison, the filling factor for AB Dor typically increases when adding spots to the ZDI original map, suggesting that the scale of the spot field is comparable to or greater than that from the ZDI map alone.

\subsection{X-ray  rotational modulation} 

This change in the extent and location of the X-ray bright regions is likely to be apparent in the degree to which the emission measure changes as the star rotates and bright regions pass behind the star. 
The plot of Fig.~\ref{005rotmod}  shows the X-ray rotational modulation of V374 Peg after adding to the ZDI map the large scale spots  (brightness level above 0.05) and a magnetic  intensity of  1~kG.  The curves are  normalized to their maximum values, thus they show the influence of the spots on the rotational modulation of the emission measure.
For the phases from 0 to 0.5, the ``new'' curves follow  the curve calculated from the ZDI naked map. For the phases  0.5 to 1 the difference between the ``original'' curve and the new curves is accentuated and is more and more visible with larger spots and stronger field strengths. 

\begin{figure*}
\resizebox{18.5cm}{!}{
\includegraphics{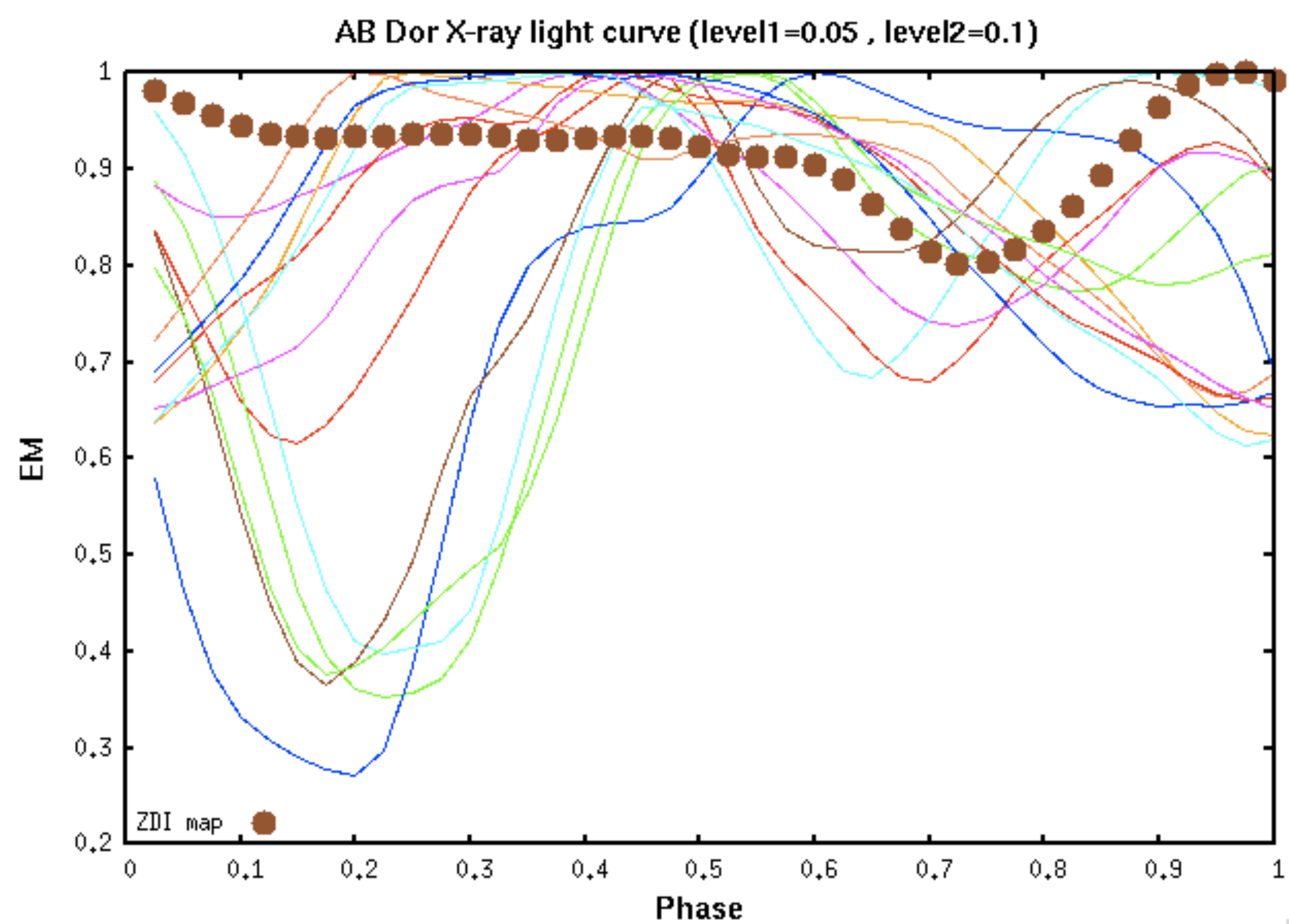}
\includegraphics{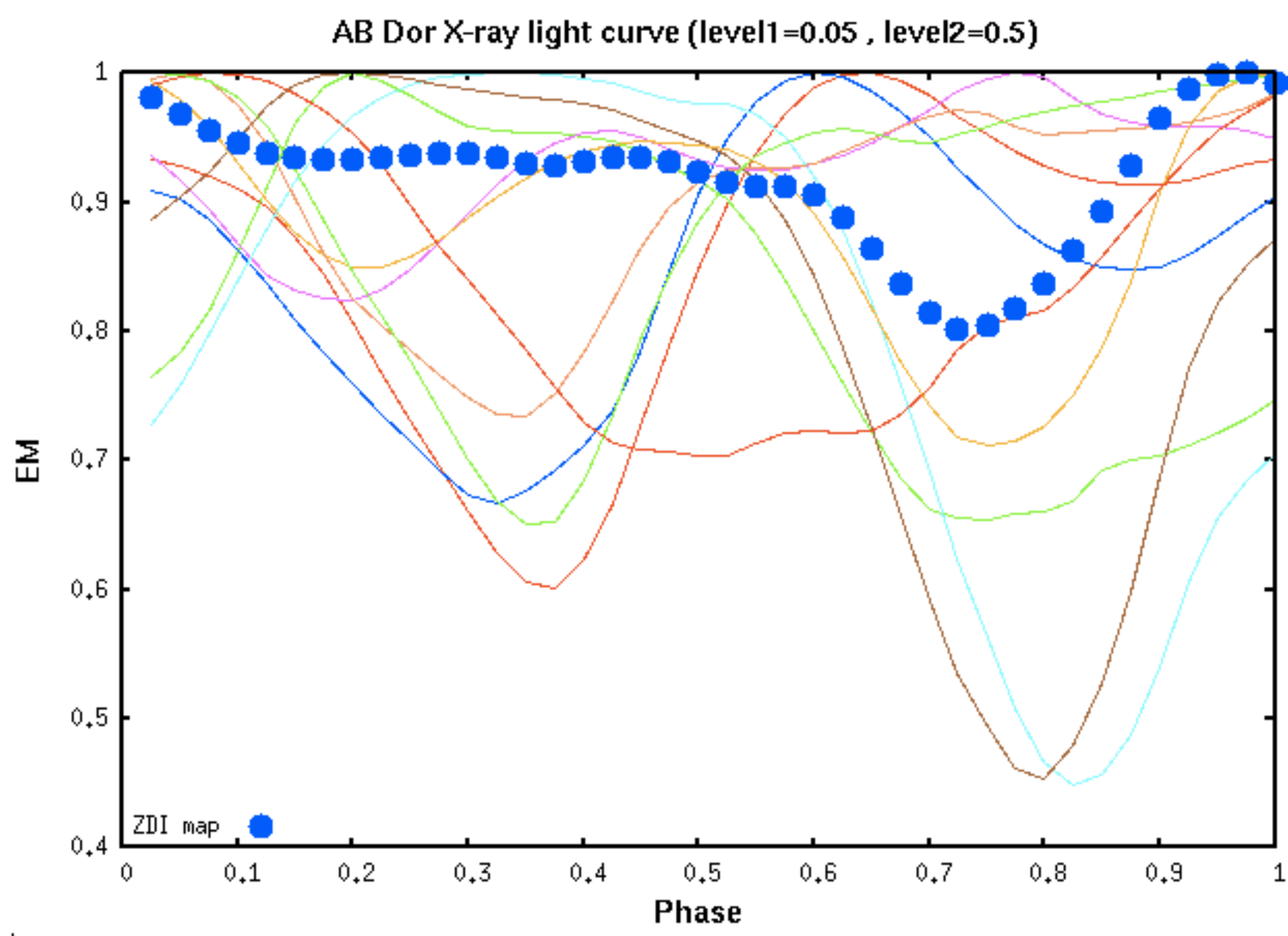} 
\includegraphics{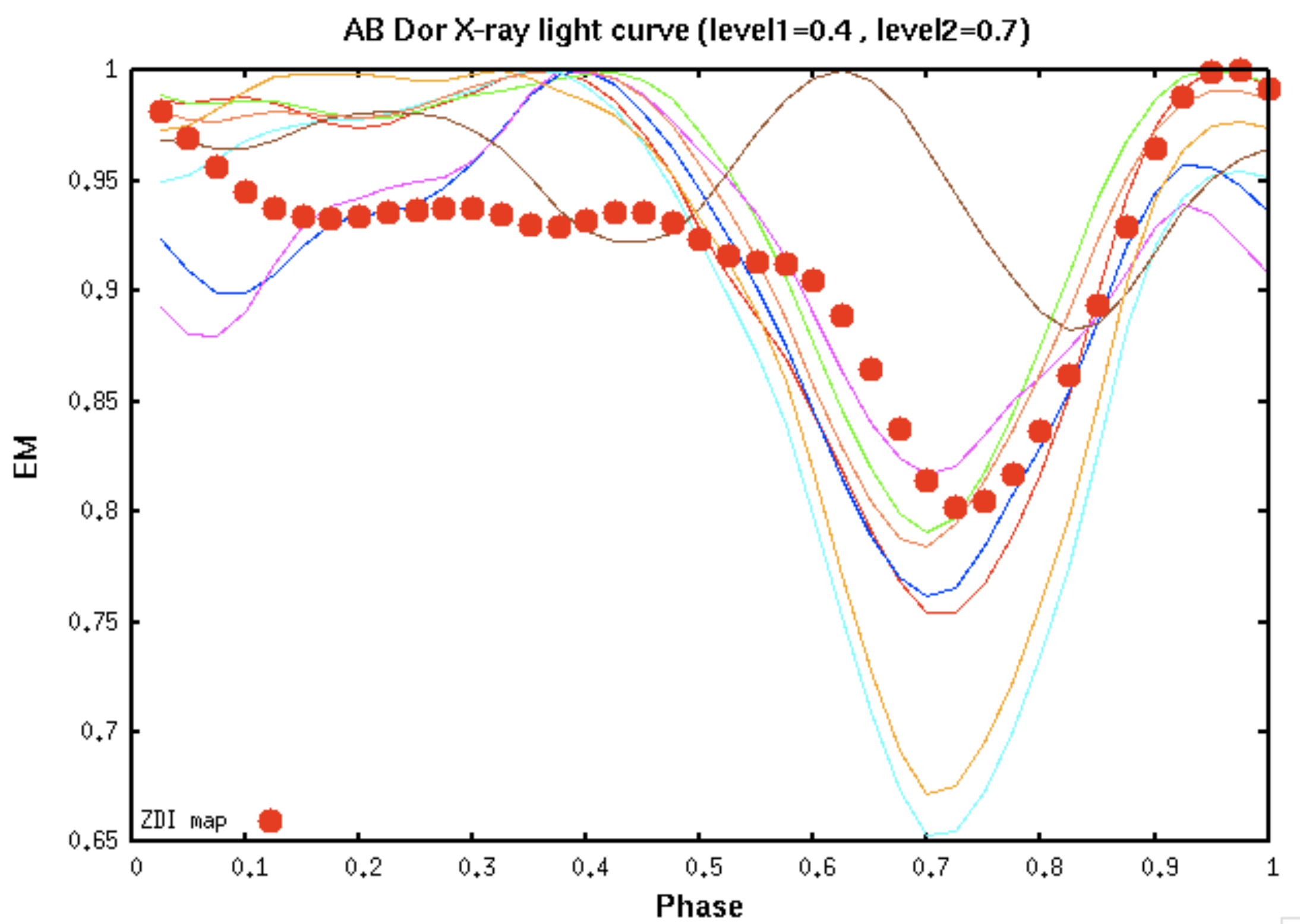}}
\caption{X-ray rotational modulation corresponding to  the 3 different configuration of spots for AB Dor (case 1 on the left, case 2 in the middle and case 3 on the right). The dots represent the X-ray rotational modulation calculated from the  ZDI map. The lines represent the  X-ray rotational modulation  using  the created maps (ZDI + spot maps). }
\label{ABDorRotmod}
\end{figure*}

In the case of AB Dor, the nature of the spot field makes a significant difference to the shape of the rotational modulation. It is clear from Fig.~\ref{ABDorRotmod} that not only the amplitude of the rotational modulation, but also the phases of the major maxima and minima can be dramatically altered when the spot field is added. 




\section{Discussion and Conclusion}

 Thanks to Zeeman-Doppler Imaging it is now possible to reconstruct the magnetic field distribution on stellar surfaces. This technique exploits {\revised the properties of the Zeeman effect on circular polarization in spectral lines (sensitive to the field strength and orientation),} as well as the broadening of the spectral lines due to the Doppler effect as the star rotates. This method complements traditional Doppler Imaging of stellar surface brightness distributions, which maps the dark starspots on stellar surfaces.  Since the Zeeman signature is suppressed in the dark regions of the stellar surface, magnetograms reconstructed using Zeeman-Doppler Imaging are censored in that they do not reconstruct reliably the field in the starspots. These two methods, therefore, are sensitive to different aspects of the stellar magnetic field. Doppler Imaging provides information about the number and distribution of dark spots on the stellar surface, while Zeeman-Doppler Imaging provides information about the distribution and (crucially) the polarity of the magnetic field in the bright regions. 
 
 We can observe the interaction of these two aspects of the magnetic field on the Sun. At the start of a new cycle, there are few spots and the large scale field is well characterised by a simple North-South dipole aligned with the rotation axis. Over the course of the solar cycle, magnetic flux emerges through the surface in a predominantly East-West orientation and is acted on by differential rotation, meridional flow and diffusion. The net drift is towards the poles, and it is the build-up of flux at the poles that eventually brings about the reversal of the large-scale field that characterises the magnetic cycle. The addition of progressively more and more azimuthal field through the cycle also causes the drift of the axis of the large-scale field from a predominantly North-South orientation at cycle minimum towards the equator and eventually to a complete reversal. This large scale field can be observed in eclipses and in the interplanetary medium. Its variation over the course of the cycle is a direct consequence of the variation of the flux that emerges through the solar surface to be observed as sunspots.
 
 The Sun is, however, a relatively inactive star, with low spot coverage. Many of the stars that have been (Zeeman) Doppler imaged are much more active, with a larger coverage of spots and a greater total flux. The distribution of spots may not be confined to discrete latitudes as on the Sun, but may extend across the whole surface, often with a large, heavily-spotted region at the pole. Mixed polarity flux may also extend across all latitudes. It is these differences that have prompted our present study of the influence of the spot field on the global structure of the coronal magnetic field that may be inferred from Zeeman-Doppler maps alone.
 
 
 

We have therefore created spot-maps of two stars: V374 Peg which is a low mass fully convective  star with a simple, strong dipolar  magnetic field and weak, low-latitude spots \citep{Morin.2008b} and AB Dor,  a solar mass star with a rather complicated magnetic field topology and extensive spot coverage, especially at the pole \citep{Donati.1999}. Three different spot scales  were used  to create several V374 Peg spot-maps, each spot  with 500~G or  1~kG  of magnetic intensity.    AB Dor's spot-maps were also created using different scales to define the dimension of the spots and two magnetic intensities were given for the spots (refer to section 3 for more details). The polarity of each spot is randomly allocated and by reproducing many times the spot-maps with random polarity it has been possible to present  a statistical distribution of  coronal parameters. We note that we have used only a simple model of the spot magnetic flux to illustrate the nature of its effect. A more sophisticated treatment would allow a profile of magnetic field strength across the spot, varying spot maximum field strengths and the possibility of mixed field polarities within large spots.

 We find  that after adding  the magnetic field from the spots to that from the ZDI magnetogram,  the corona of both stars seems to be affected. The magnitude of this effect increases as the flux in the spots increases. 
 V374 Peg's added spots are small (compared to the total surface area) and have simple shapes.  They add some complexity to the overall field, creating small magnetic structures (and hence decreasing of the magnetic  field  that crosses  the equator). Their inclusion typically increases the total magnetic flux as well as the parameters that are derived from it: the plasma pressure and  density and hence the X-ray emission measure. The tilt angle of the dipole component varies by 2 or 3 degrees around the initial value and is related to the polarity of the added spots. 
 
AB Dor's field is more complex and the distribution of spots at its surface has a greater effect. There is a large spot at the visible pole of the star which, since it is assigned a single (randomly-chosen) polarity, produces large scale magnetic field. Many of these are open and therefore the inclusion of the spot field increases the fraction of open and therefore wind-bearing flux. The closed magnetic flux  from this region tend to run principally North-South and so the dipole axis (initially at $32^{\circ}$) tends to shift closer to the rotation axis when the spot field is included. 

The X-ray rotational modulation  has also been  reconstructed for each star using the created spots maps and compared to the original X-ray light curve. As the added spots in the case of V374 Peg are relatively small their effect on the X-ray light curve  is only  significant at phases between 0.7 and 0.9 (cf. Fig.~\ref{005rotmod}). By comparison, the X-ray light curves of AB Dor derived from the spot-maps are very different in amplitude and shape  from the initial one.   

The work presented in this paper shows  that the coronal properties inferred from Zeeman-Doppler maps alone can be significantly affected by the inclusion of magnetic field in the spots. The magnitude of this effect depends both on the relative amount of flux in the dark and bright regions of the stellar surface, and also on the distribution of spots. Low-latitude spots, arranged closely together, tend to add small-scale, East-West field. This increases the magnitude of the X-ray emission measure while decreasing its filling factor. A large polar spot, however, typically makes the overall field closer to an aligned dipole, and increases both the amount of open flux and the X-ray emission measure.

It seems, then, that for a star with a small coverage of spots, such as V374 Peg, that the neglect of the field in the spots has only a modest effect on the coronal geometry. For a star such as AB Dor, however, the spot coverage is much greater. Depending on the magnitude of the magnetic flux hidden in these spots, their neglect may have a significant effect on the coronal structure, and in particular on the magnitude and rotational modulation of the X-ray emission measure.




 \bibliographystyle{mn2e}
 \bibliography{bib}

\begin{thebibliography}{15}
\expandafter\ifx\csname natexlab\endcsname\relax\def\natexlab#1{#1}\fi

\bibitem[{{Altschuler} \& {Newkirk}(1969)}]{Altschuler.1969}
{Altschuler} M.~D., {Newkirk} G., 1969, \solphys, 9, 131

\bibitem[{{Berdyugina}(2005)}]{Berdyugina.2005}
{Berdyugina} S.~V., 2005, Living Reviews in Solar Physics, 2, 8

\bibitem[{{Chandrasekhar}(1961)}]{Chandrasekhar.1961}
{Chandrasekhar} S., 1961, {Hydrodynamic and hydromagnetic stability},
  {Chandrasekhar, S.}, ed.

\bibitem[{{Donati} \& {Collier Cameron}(1997)}]{Donati.Cameron.1997}
{Donati} J.-F., {Collier Cameron} A., 1997, \mnras, 291, 1

\bibitem[{{Donati} {et~al.}(1999){Donati}, {Collier Cameron}, {Hussain}, \&
  {Semel}}]{Donati.1999}
{Donati} J.-F., {Collier Cameron} A., {Hussain} G.~A.~J., {Semel} M., 1999,
  \mnras, 302, 437

\bibitem[{{Donati} {et~al.}(2006){Donati}, {Forveille}, {Cameron}, {Barnes},
  {Delfosse}, {Jardin}, \& {Valenti}}]{Donati.2006b}
{Donati} J.-F., {Forveille} T., {Cameron} A.~C., {Barnes} J.~R., {Delfosse} X.,
  {Jardin} M.~M., {Valenti} J.~A., 2006, Science, 311, 633

\bibitem[{{Donati} {et~al.}(2009){Donati}, {Morin}, {Delfosse}, {Forveille},
  {Far{\`e}s}, {Moutou}, \& {Jardine}}]{Donati.2009}
{Donati} J.-F., {Morin} J., {Delfosse} X., {Forveille} T., {Far{\`e}s} R.,
  {Moutou} C., {Jardine} M., 2009, in American Institute of Physics Conference
  Series, Vol. 1094, American Institute of Physics Conference Series,
  {Stempels} E., ed., pp. 130--139

\bibitem[{{Donati} {et~al.}(1997){Donati}, {Semel}, {Carter}, {Rees}, \&
  {Collier Cameron}}]{Donati.1997}
{Donati} J.-F., {Semel} M., {Carter} B.~D., {Rees} D.~E., {Collier Cameron} A.,
  1997, \mnras, 291, 658

\bibitem[{{Hale}(1908)}]{Hale.1908}
{Hale} G.~E., 1908, \apj, 28, 315

\bibitem[{{Hussain} {et~al.}(2007){Hussain}, {Jardine}, {Donati}, {Brickhouse},
  {Dunstone}, {Wood}, {Dupree}, {Collier Cameron}, \& {Favata}}]{Hussain.2007}
{Hussain} G.~A.~J., {Jardine} M., {Donati} J.-F., {Brickhouse} N.~S.,
  {Dunstone} N.~J., {Wood} K., {Dupree} A.~K., {Collier Cameron} A., {Favata}
  F., 2007, \mnras, 377, 1488

\bibitem[{{Jardine} {et~al.}(2006){Jardine}, {Cameron}, {Donati}, {Gregory}, \&
  {Wood}}]{Jardine.2006}
{Jardine} M., {Cameron} A.~C., {Donati} J.-F., {Gregory} S.~G., {Wood} K.,
  2006, \mnras, 367, 917

\bibitem[{{Johnstone} {et~al.}(2010){Johnstone}, {Jardine}, \&
  {Mackay}}]{Johnstone.2010}
{Johnstone} C., {Jardine} M., {Mackay} D.~H., 2010, \mnras, 404, 101

\bibitem[{{Morin} {et~al.}(2008){Morin}, {Donati}, {Forveille}, {Delfosse},
  {Dobler}, {Petit}, {Jardine}, {Cameron}, {Albert}, {Manset}, {Dintrans},
  {Chabrier}, \& {Valenti}}]{Morin.2008b}
{Morin} J., {Donati} J.-F., {Forveille} T., {Delfosse} X., {Dobler} W., {Petit}
  P., {Jardine} M.~M., {Cameron} A.~C., {Albert} L., {Manset} N., {Dintrans}
  B., {Chabrier} G., {Valenti} J.~A., 2008, \mnras, 384, 77

\bibitem[{{van Ballegooijen} {et~al.}(1998){van Ballegooijen}, {Cartledge}, \&
  {Priest}}]{vanBallegooijen.1998}
{van Ballegooijen} A.~A., {Cartledge} N.~P., {Priest} E.~R., 1998, \apj, 501,
  866

\bibitem[{{Willis} \& {Young}(1987)}]{Willis.1987}
{Willis} D.~M., {Young} L.~R., 1987, Geophysical Journal International, 89,
  1011

\end{thebibliography}

\appendix

\end{document}